\documentclass[aps,prc,twocolumn,floatfix,superscriptaddress]{revtex4}
\usepackage{longtable}
\usepackage{graphicx}
\usepackage{dcolumn}
\usepackage{color}
\usepackage{xcolor}
\usepackage{float}
\usepackage{amsmath}
\usepackage{soul}

\begin{document}

\title{\boldmath A~New~Evidence of Interplay Between Tetrahedral and Octahedral Symmetries and~Symmetry Breaking: Exotic Rotational Bands in $^{152}$Sm}
\author{S.~Basak}
\affiliation{Variable Energy Cyclotron Centre, Kolkata - 700 064, India}
\affiliation{Homi Bhabha National Institute, Training School Complex, Anushakti Nagar, Mumbai - 400 094, India}
\author{D.~Kumar}
\affiliation{Variable Energy Cyclotron Centre, Kolkata - 700 064, India}
\affiliation{Homi Bhabha National Institute, Training School Complex, Anushakti Nagar, Mumbai - 400 094, India}
\author{T.~Bhattacharjee}
\thanks{Corresponding author}
\email{btumpa@vecc.gov.in}
\affiliation{Variable Energy Cyclotron Centre, Kolkata - 700 064, India}
\affiliation{Homi Bhabha National Institute, Training School Complex, Anushakti Nagar, Mumbai - 400 094, India}
\author{I.~Dedes}
\affiliation{Institute of Nuclear Physics, Polish Academy of Sciences, PL-31 342 Krak\'ow, Poland}
\author{J.~Dudek}
\affiliation{Universit\'e de Strasbourg, CNRS, IPHC UMR 7178, F-67 000 Strasbourg, France}
\affiliation{Institute of Nuclear Physics, Polish Academy of Sciences, PL-31 342 Krak\'ow, Poland}
\affiliation{Institute of Physics, Marie Curie-Sk\l odowska University, PL-20 031 Lublin, Poland}
\author{A.~Pal}
\affiliation{Variable Energy Cyclotron Centre, Kolkata - 700 064, India}
\affiliation{Homi Bhabha National Institute, Training School Complex, Anushakti Nagar, Mumbai - 400 094, India}
\author{S.~S.~Alam}
\affiliation{Government General Degree College, Chapra - 741 123, West Bengal, India}
\author{A.~Saha}
\affiliation{ICFAI University Tripura, Agartala, Tripura-799 210, India}
\author{A.~K.~Sikdar}
\affiliation{Variable Energy Cyclotron Centre, Kolkata - 700 064, India}
\affiliation{Homi Bhabha National Institute, Training School Complex, Anushakti Nagar, Mumbai - 400 094, India}
\author{J.~Nandi}
\affiliation{Variable Energy Cyclotron Centre, Kolkata - 700 064, India}
\affiliation{Homi Bhabha National Institute, Training School Complex, Anushakti Nagar, Mumbai - 400 094, India}
\author{Shabir~Dar}
\altaffiliation [Present address:~] {Department of Physics and Astronomy, Division of Applied Nuclear Physics, Uppsala University, 75120 Uppsala, Sweden}
\affiliation{Variable Energy Cyclotron Centre, Kolkata - 700 064, India}
\affiliation{Homi Bhabha National Institute, Training School Complex, Anushakti Nagar, Mumbai - 400 094, India}
\author{A.~Baran}
\affiliation{Institute of Physics, Marie Curie-Sk\l odowska University, PL-20 031 Lublin, Poland}
\author{A.~Gaamouci}
\affiliation{Institute of Nuclear Physics, Polish Academy of Sciences, PL-31 342 Krak\'ow, Poland}
\author{D.~Rouvel}
\affiliation{Universit\'e de Strasbourg, CNRS, IPHC UMR 7178, F-67 000 Strasbourg, France}
\affiliation{Lyc\'ee Kl\'eber, 25 place de Bordeaux, F-67 000, Strasbourg, France}
\author{S.~Samanta}
\affiliation{UGC-DAE CSR, Kolkata Centre, Kolkata - 700 098, India}
\author{S.~Chatterjee}
\affiliation{UGC-DAE CSR, Kolkata Centre, Kolkata - 700 098, India}
\author{R.~Raut}
\affiliation{UGC-DAE CSR, Kolkata Centre, Kolkata - 700 098, India}
\author{S.~S. Ghugre}
\affiliation{UGC-DAE CSR, Kolkata Centre, Kolkata - 700 098, India}
\author{A.~Adhikari} 
\affiliation{Indian Institute of Engineering Science and Technology, Howrah-711 103, India}
\author{Y.~Sapkota}
\altaffiliation [Present address:~] {Department of Physics, Dudhnoi College, Dudhnoi, Goalpara, Assam-783120, India}
\affiliation{{Assam Don Bosco University, Tapesia Gardens, Sonapur, Assam- 782 402, India}}
\author{R.~Rahaman}
\affiliation{Indian Institute of Engineering Science and Technology, Howrah-711 103, India}
\author{Ananya~Das}
\affiliation{{Dream Institute of Technology, Samali, Kolkata -  700 104}}
\author{A.~Gupta}
\affiliation{Indian Institute of Engineering Science and Technology, Howrah-711 103, India}
\author{A.~Bisoi} 
\affiliation{Indian Institute of Engineering Science and Technology, Howrah-711 103, India}
\author{S.~Sharma}
%\altaffiliation [Present address:~] {Manipal~Institute~of~Technology, Manipal~Academy~of~Higher~Education, Manipal-576104, Karnataka, India}
\affiliation{{Manipal University Jaipur, Jaipur-303007, Rajasthan, India}}
\author{S.~Das}
%\altaffiliation [Present address:] {Brainware University, Kolkata-700125, India}
\affiliation{Indian Institute of Engineering Science and Technology, Howrah-711 103, India}
\author{A.~Bhattacharyya}
\affiliation{Saha Institute of Nuclear Physics, 1/AF Bidhannagar, Kolkata-700 064, India}
\author{P.~Das}
\affiliation{Saha Institute of Nuclear Physics, 1/AF Bidhannagar, Kolkata-700 064, India}
\author{U.~Datta}
\affiliation{Saha Institute of Nuclear Physics, 1/AF Bidhannagar, Kolkata-700 064, India}
%\author{S.~Chattopadhyay}
%\affiliation{Saha Institute of Nuclear Physics, 1/AF Bidhannagar, Kolkata-700 064, India}
\author{I.~Ray}
\affiliation{Jadavpur University, Kolkata-700 032, India}
\author{J.~Yang}
\affiliation{School of Physics and Electronic Technology, Liaoning Normal University, Dalian, Liaoning Province, 116 029, P.R.~China}
\author{D.~Curien}
\affiliation{Universit\'e de Strasbourg, CNRS, IPHC UMR 7178, F-67 000 Strasbourg, France}
\author{G.~Duch\^{e}ne}
\affiliation{Universit\'e de Strasbourg, CNRS, IPHC UMR 7178, F-67 000 Strasbourg, France}

\date{\today}

\begin{abstract}

We report on an experimental evidence for a new, second tetrahedral band in $^{152}_{\;\;62}$Sm$^{}_{90}$. It was populated via fusion evaporation reaction, $^{150}{\rm Nd}(\alpha, 2n)^{152}$Sm, employing 26 MeV beam of $\alpha$ particles from K-130 cyclotron at Variable Energy Cyclotron Centre, Kolkata, India. The newly observed possible mixed parity sequence with absence of E2 and strong indication of E3 transitions is consistent with the spectroscopic criteria for a tetrahedral-symmetry rotational band that could be constructed from the allowed spin-parity assignments. This structure differs from the structure of the band previously found in the same nucleus, the new one manifesting tetrahedral symmetry not accompanied by the octahedral one. Our new experimental results are interpreted in terms of group representation theory and collective nuclear-motion theory of Bohr. We propose to generalize the notion of the tetrahedral vibrational bands and believe that our new experimental results support a number of theory predictions related to nuclear tetrahedral symmetry published earlier and bring a new light into the issue of spontaneous symmetry breaking in heavy nuclei.

\end{abstract}

\maketitle

\section{Introduction}
\label{intro}

Notions of nuclear shapes can be naturally associated with geometrical symmetries,  described with the help of realistic phenomenological mean-field theory and group theory, for the latter one, see Refs.~\cite{MMH62,GFK63,JFC97}. The so called double tetrahedral point group, T$^{\rm D}_{\rm d}$, is needed to describe the symmetries of the systems of Fermions like atomic nuclei. It possesses 4-dimensional irreducible representations, what implies an existence of exotic 4-fold degeneracies of nucleonic levels in contrast to double (the so-called Kramers degeneracies) present in all other non-spherical nuclei. Large scale nuclear mean-field calculations show that tetrahedral symmetry induces strong gap-openings at tetrahedral magic numbers $Z, N=32$, 40, 56, 70, 90, 112, 136 in the single-nucleon energy-spectra, see Refs.~\cite{XLi94,JDu88,JDu06}.

Let us remind the reader at this point that one of the most convenient ways of describing the nuclear surfaces is to use the basis of spherical harmonics $\{Y_{\lambda \mu} (\vartheta,\varphi)\}$ with the index $\lambda$ called multipolarity varying typically between $\lambda=2$ (quadrupole deformation) and the multipolarity cutoff parameter $\lambda_{\rm max}$. With these prerequisites the expression of the nuclear surface, say $\Sigma$, takes the form 
\begin{equation}
  \Sigma: R(\vartheta,\varphi) = R_o c(\alpha) [1 + \sum_{\lambda=2}^{\lambda_{\rm max}} \alpha_{\lambda, \mu} Y_{\lambda,\mu}(\vartheta,\varphi)],
                                                                  \label{Eqn_01}
\end{equation}
where by definition $R_o=r_o * A^{1/3}$ with, typically, $r_o=1.2$~fm and $A$ denotes the mass number, whereas the function $c(\alpha)$ assures the nuclear constant volume condition, and the abbreviation $\alpha \equiv \{\alpha_{\lambda,\mu}\}$. We refer to the symmetries implied by quadrupole deformations, $\alpha_{20}$ and $\alpha_{22}$ and octupole pear-shape deformation $\alpha_{30}$ as ``traditional'' and to all the others as exotic. In this article tetrahedral (also called ``pyramid like'') and octahedral (``diamond like'') deformations -- and the corresponding exotic symmetries will be of central interest.

One can demonstrate that the tetrahedral symmetry surfaces can be expressed using  only specific selection of spherical harmonics. Following Ref.~\cite{JDu07}, tetrahedral symmetry can be described according to an increasing multipole order with the help of
\begin{equation}
   \lambda=3:\quad \alpha_{3,\pm 2} \equiv t_1,
                                                                  \label{Eqn_02}
\end{equation}
there are no solutions possible up to multipolarity 
\begin{equation}
   \lambda=7:\quad \alpha_{7,\pm 2} \equiv t_2,\;\;{\rm and}\;\; 
                   \alpha_{7,\pm 6} \equiv -\sqrt{11/13}\, t_2,
                                                                  \label{Eqn_03}
\end{equation}
whereas the higher orders correspond to odd multipolarities $\lambda > 7$. Similar considerations for octahedral symmetry show that the lowest order solutions correspond to multipolarity 
\begin{equation}
   \lambda=4:\quad \alpha_{4, 0} \equiv o_1,\;\;{\rm and}\;\; 
                   \alpha_{4,\pm 4} = -\sqrt{5/14}\, o_1
                                                                  \label{Eqn_04}
\end{equation}
whereas higher multipolarity solutions are possible for even $\lambda$ only, beginning with
\begin{equation}
   \lambda=6:\quad \alpha_{6, 0} \equiv o_2,\;\;{\rm and}\;\; 
                   \alpha_{6,\pm 4} = -\sqrt{11/13}\, o_2.
                                                                  \label{Eqn_05}
\end{equation}

Having determined how to introduce mathematically the new exotic symmetries of interest one needs to chose the theoretical approach of calculating the corresponding total nuclear potential energies (in our case microscopic-macroscopic method (MMM)) and open the way of experimental identification of the symmetry predictions.
Tagami et al., Ref.~\cite{STa13}, employing group representation theory predict  spin-parity structures of rotational bands built on $I^\pi=0^+$ band-heads in T$_{\rm d}$ symmetric nuclei, which according to the associated A$_{1}$ irreducible representation is composed of the following $I^\pi$ states:
\begin{equation}
\!\!\!\! \underbrace{{\rm A_1}:\, 
  0^+,3^-,4^+,
  \underbrace{(6^+,6^-)}_{\rm doublet},7^-,8^+,
  \underbrace{(9^+,9^-)}_{\rm doublet},
  \underbrace{(10^+,10^-)}_{\rm doublet},\cdots}
           _{{A_1:~Parabolic~} E-{\rm vs.}- I~{\rm sequence ~built~on}~I^\pi=0^+~{\rm band~head}}
                                                                  \label{Eqn_06}
\end{equation}
satisfying, to a good approximation, the quantum rotor relation
\begin{equation}
     E_I \propto \frac{\hbar^2}{2\mathcal{J}_{T_{\rm d}}} I(I+1) ,
                                                                  \label{Eqn_07}
\end{equation}
with an effective moment of inertia $\mathcal{J}_{T_{\rm d}}$. The authors of Ref.~\cite{STa13} employed  microscopic, realistic, spin and parity projected, constrained  Hartree-Fock approach, reproducing  relation~(\ref{Eqn_06}) even though the underlying computer programs contained neither symmetry nor group theory information. The only impact of symmetry was implicit; the authors employed the T$_{\rm d}$-symmetry constraint by imposing the values of the multipole moment~$Q_{32}$.

Let us emphasize the presence in Eq.~(\ref{Eqn_06}) of both even and odd spin members of both parities. Some states appear degenerate, such as I$^\pi=6^\pm, 9^\pm, 10^\pm$ etc., some spins (I$=1,2$ and 5) are totally missing. According to group-theory nomenclature, structures built on irreducible representations $A_2$, $E$, $T_1$ and $T_2$, represent the excited T$_{\rm d}$ bands built on the band-heads I$^\pi=0^-,2^\pm, 1^+$ and $1^-$, respectively; they will not be discussed in the present article. The first experimental evidence of a tetrahedral symmetry in a medium heavy (non-cluster) nucleus $^{152}$Sm,  was found earlier, and published by other authors, Ref.~\cite{prc2018}, who emphasized the coexistence between tetrahedral and octahedral, T$_{\rm d}$ and  O$_{\rm h}$, symmetries. In the following we refer to the corresponding rotational band of Ref.~\cite{prc2018} as T$_{\rm d}(1)$. 

It turns out that the T$_{\rm d}(1)$ band members related to the irreducible representation $A_1$, cf.~Eq.~(\ref{Eqn_06}), can be found in irreducible representations $A_{1g}$ and $A_{2u}$ according to exact octahedral symmetry O$_{\rm h}$ (let us recall at this point that T$_{\rm d}$ is a sub-group of O$_{\rm h}$):
\begin{equation}
     \underbrace{
      A_{1g}: \;  0^+,4^+,6^+,8^+,9^+,10^+,\; \ldots\,,\;\;I^\pi \leftrightarrow I^+
     }_{{\rm Exact~O_{\rm h}}~I^\pi = 0^+~{\rm band-head:~A~certain~parabolic~sequence}} \!\!\!\!,
                                                               \label{Eqn_08}
\end{equation}
and
\begin{equation}
     \underbrace{
      A_{2u}:\; 3^-,6^-,7^-,9^-,10^-,11^-, \ldots\,,\;I^\pi \leftrightarrow I^-
     }_{{\rm Exact~O_{\rm h}}~I^\pi=3^-~{\rm band-head:~Another~parabolic~sequence}} \!\!,
                                                               \label{Eqn_09}
\end{equation}
with the associated effective moments of inertia  $\mathcal{J}_{A_{1g}}$ and $\mathcal{J}_{A_{2u}}$.
According to group theory, irreducible representations span independent sets of states so that {\em a priori} no common features among the parabolic sequences in Eqs.~(\ref{Eqn_08}) and (\ref{Eqn_09}) are to be expected, and in particular, no correlation between $\mathcal{J}_{A_{1g}}$ and $\mathcal{J}_{A_{2u}}$. Yet, empirically one finds that $\mathcal{J}_{A_{1g}} \approx \mathcal{J}_{A_{2u}} \approx\mathcal{J}_{A_1}$. Since the exact O$_{\rm h}$ symmetry imposes that  $A_{1g}$ and $A_{2u}$ should be uncorrelated,  an approximate equality relation signifies that O$_{\rm h}$-symmetry is broken (here by T$_{\rm d}$-symmetry, the one which imposes that the considered moment of inertia should be approximately equal as implied by Eq.~(\ref{Eqn_06})).

One can demonstrate using standard arguments of quantum mechanics, Ref.~\cite{JDEPJ}, that for exact tetrahedral symmetry,   collective $E1$ and $E2$ transitions vanish. The absence of usually strong $E1$ and $E2$ transitions makes it difficult both to populate the corresponding states and to observe their decay via $\gamma$-detection. Consequently, identification of tetrahedral bands remains an experimental challenge, see~Refs.~\cite{MJe10,RAB10,ASa19}. 

\section{Experimental Details}
\label{expt}

The excited states of $^{152}$Sm nuclei were populated using $^{150}$Nd$\rm (^4He,2n)^{152}$Sm reaction at $E_{\rm beam} = 26$~MeV. The fusion-evaporation cross sections, $\sigma_{fe}$, is peaked around $10~\hbar$, with dominating component for $^{152}$Sm ($\sigma_{fe}\approx 1070$~mb) together with $^{151}$Sm ($\sigma_{fe}\approx 75$~mb) from ($\alpha$,3n) channel and $^{153}$Sm ($\sigma_{fe}\approx 1.65$~mb) from ($\alpha$,n) channel. Therefore $\alpha$-beam facilitated selective population of non-yrast low spin states with a lower background from yrast $\gamma$-lines. This helped observing very weak non-yrast tetrahedral structures in $^{152}$Sm, as discussed in the following Sections. De-exciting $\gamma$-transitions were detected using twelve BGO Anti Compton Suppressed (ACS) Clover HPGe detectors arranged in three rings (6 detectors at $90^{\circ}$, $3$ at $40^{\circ}$ and another 3 at $125^{\circ}$). 

Enriched $^{150}$Nd target of $\approx 10~\rm mg/cm^2$ thickness was prepared on mylar backing, using centrifuge technique. The list mode data, consisting of about $2 \times 10^9$ $\gamma$-$\gamma$ coincidence events, were collected using a digital Data Acquisition (DAQ) facility based on Pixie-16 digitizers setting a trigger condition of {Compton suppressed clover fold $\geq 1$}. The gain-matched and time-merged data were sorted using IUCPIX package of Ref.~\cite{iuc}, forming symmetric and asymmetric $\gamma$-$\gamma$ matrices and triple $\gamma$ cubes. They were employed to analyze coincidence relationships, DCO (Directional Correlation of Oriented States) ratios~\cite{kfdco} and IPDCO (Integrated Polarization Directional Correlation of Oriented States)~\cite{jones,droste} ratios with conventional high resolution $\gamma$-spectroscopic techniques using Radware~\cite{radware} and Ingasort~\cite{ingasort} packages. Singles data were used for analysis of angular distributions of $\gamma$ transitions.

%%%%%%%%%%%%%%%%%%%%%%%%%%%%%%%%%%%%%%%%%%%%%%%%%%%%%%%%%%%%%%%%%%%%%%%%%%%%%%%%%%%%%

\section{Analysis Technique}
\label{Sec-III}

In this section we present our data analysis technique that has been used to find the candidate tetrahedral band in $^{152}$Sm nucleus, from now on referred to as T$_{\rm d}(2)$-band. 

Gated spectra from $\gamma$-$\gamma$ matrices and $\gamma$-$\gamma$-$\gamma$ cubes were analyzed in search of the new transitions. We explored the possible presence of any contaminating transitions with energies similar to those depopulating the levels of $\rm T_{\rm d}(2)$-band. However, no such transitions were found in $^{151,152,153}$Sm populated in this experiment.

\begin{figure}[ht!]
\begin{center}
\includegraphics[width=\columnwidth]{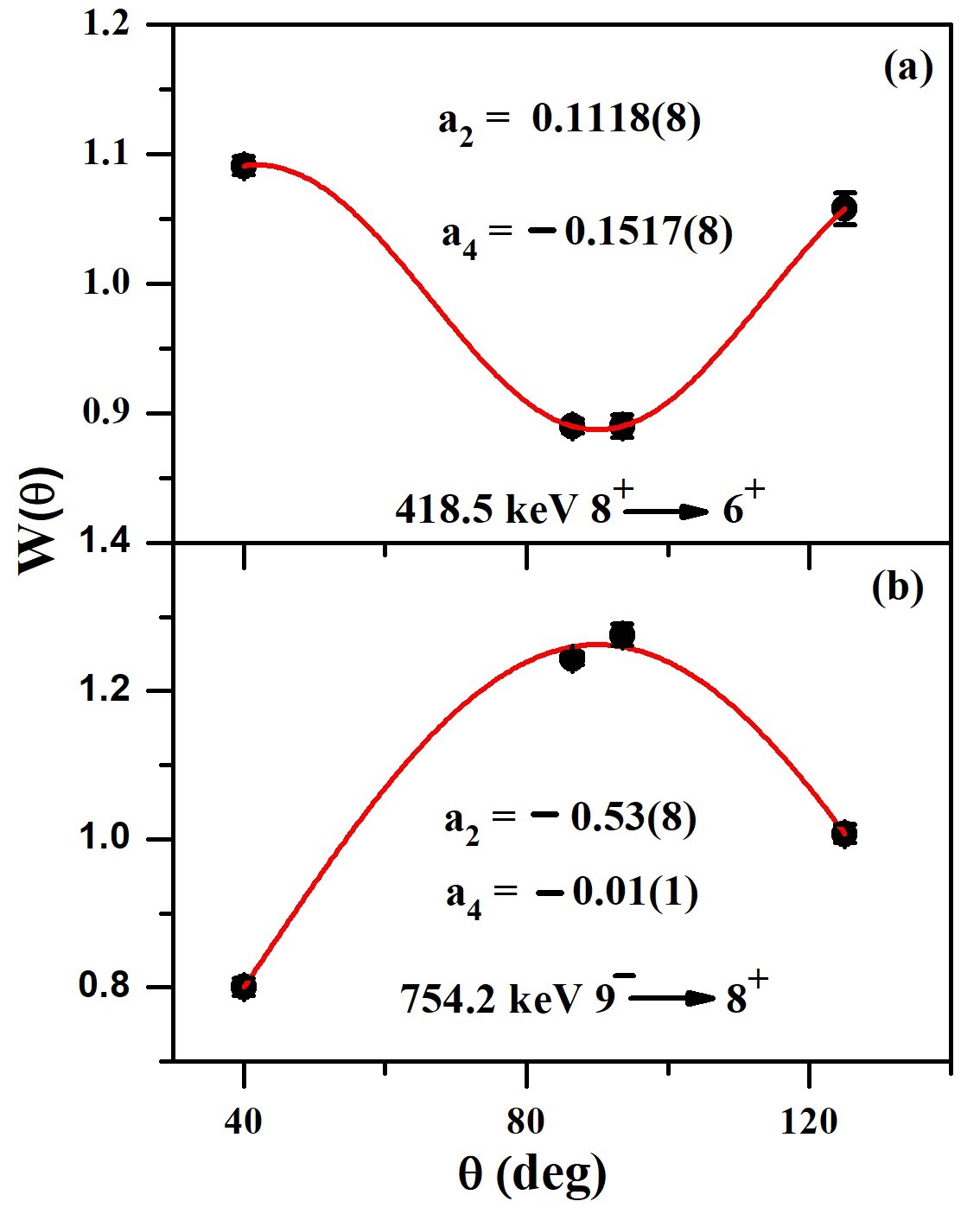}
\caption{The angular distribution (W($\theta$) vs $\theta$) fits for known quadrupole (E2) (a) and dipole (E1) (b) transitions.} 
                                                                  \label{Fig_01}
\end{center}
\end{figure}

Spin-parity identifications of observed levels were confirmed through measurements of the angular distribution $W(\theta)$, DCO ratios R$_{\rm DCO}$ and Polarization, P. W($\theta$) was determined at four angles, two of them being 86.5$^{\circ}$ and 93.5$^{\circ}$, obtained dividing the crystals of the 90$^{\circ}$ clovers into two halves, using the fact that the opening angle of the Clover detector is  7$^{\circ}$. The angular distribution coefficients ($a_2$ and $a_4$) were determined fitting $W(\theta)$-vs.-$\theta$ dependence and are shown in Fig.~\ref{Fig_01} for two known transitions, {\em viz.}, 418.5~keV (E2), panel (a), and 754.2~keV (E1), panel (b) in $^{152}$Sm,  as known from the NNDC database,~Ref.~\cite{ensdf}.  

Analysis R$_{\rm DCO}$ was performed using asymmetric matrix of 
$90^{\circ}$-vs.-$125^{\circ}$ detectors and quadrupole gates. The details of the method can be found in Ref.~\cite{ASa19}. 

The clover HPGe detectors can also be used to determine the linear polarization of $\gamma$-rays, P$_{\rm exp}$,~Ref.~\cite{jones} by observing their parallel and perpendicular scattering in the clover segments. In the present work, P$_{\rm exp}$ was measured from the polarization asymmetry $\Delta_{\rm IPDCO}$ and polarization sensitivity,  Q, Refs.~\cite{duchene1,duchene2,subhendu}. For method of IPDCO analysis, see Ref.~\cite{ASa19} employing detectors placed at $90^{\circ}$. The clover polarization sensitivity was adopted from Ref.~\cite{subhendu}, where it was determined using measured $a_2$ and $a_4$ coefficients for the type of clover detectors used in the present work. 

\begin{figure}[ht!]
\begin{center}
\includegraphics[width=\columnwidth]{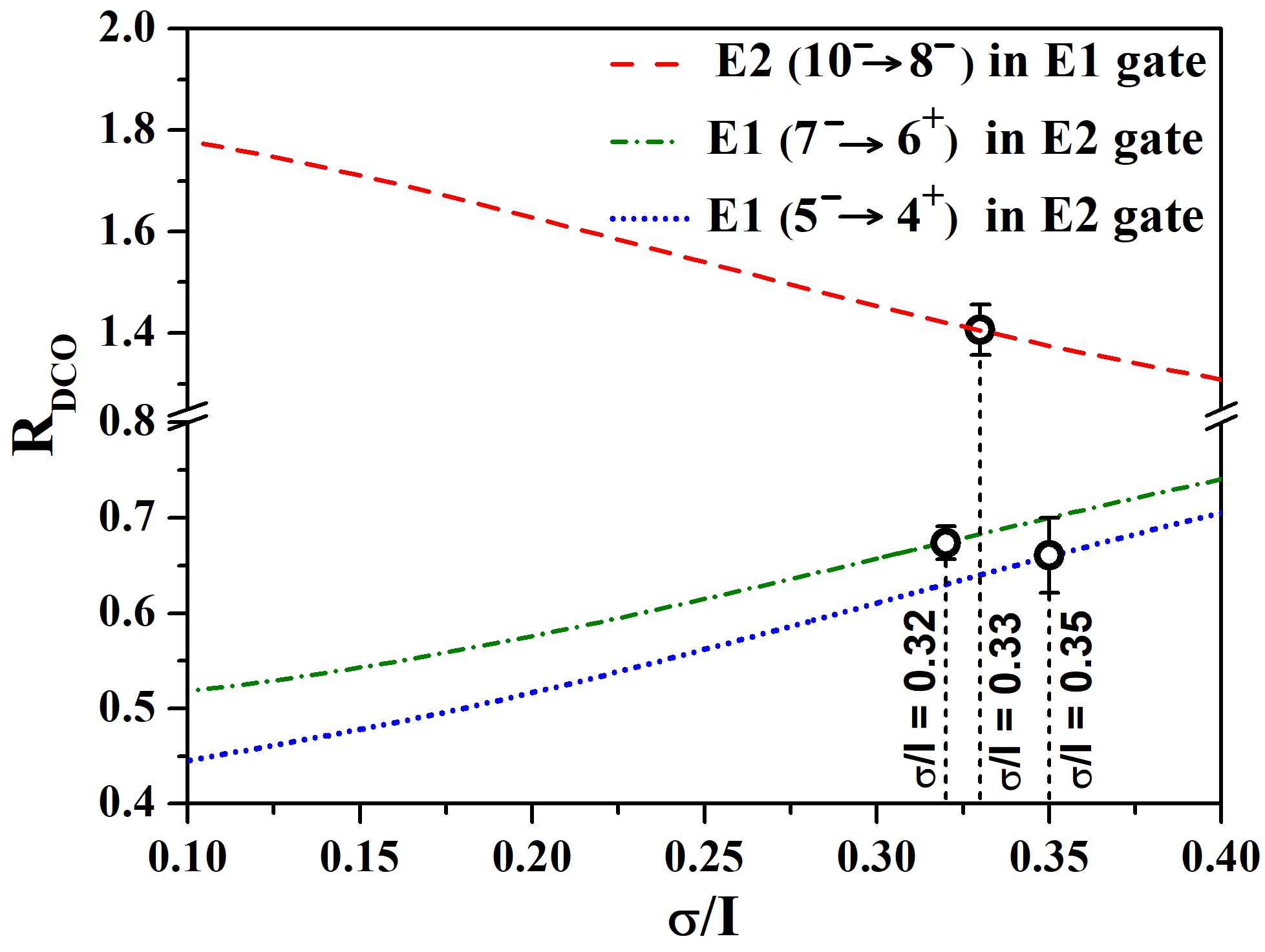}
\caption{This figure illustrates the determination of $\sigma /I$ introduced in the text. This has been done using the R$_{\rm DCO}$ values of known E1 and E2 transitions which are depopulating excited levels with known spins, as shown. The  R$_{\rm DCO}$ values (calculated from ANCORR code) as a function of $\sigma /I$ are drawn with lines as indicated in the legend. Experimental data points for three such transitions and the corresponding $\sigma/I$ required to reproduce these experimental data are shown in the figure as circles. The average of these three values is  0.33; it was used for the calculation of $a_2$, $a_4$ and R$_{\rm DCO}$ in the present work. See text for details.}

                                                                  \label{Fig_02}
\end{center}
\end{figure}

The angular distribution coefficients $a_2$ and $a_4$ and the ratios R$_{\rm DCO}$ were also calculated from ANCORR code,~Ref.~\cite{ancorr}, using partial alignment of the spin projection $m$ on the beam axis for given angular momentum $I$ represented by a Gaussian distribution. The $\sigma/I$ value was determined using the R$_{\rm DCO}$ values of known transitions and comparing with the calculated R$_{\rm DCO}$ from ANCORR code. The average of the $\sigma/I$ values determined using the known transitions was found to be 0.33, as displayed in Fig.~\ref{Fig_02}. This average value of $\sigma/I$ was used for the subsequent calculations for the new transitions placed in T$_{\rm d}(2)$-band. 

\noindent

Similarly, calculated values of linear polarization P$_{\rm calc}$ were obtained following Refs.~\cite{faghana,deng} with the help of experimentally determined $a_2$ and $a_4$, while varying the mixing ratios $\delta$ between $\delta = -20$ and $\delta = +20$. The following equation was used to determine P$_{\rm calc}$. 
\begin{equation}
   P_{calc} 
   = 
   \pm \frac{\sum _{\lambda = 2, 4} a_{\lambda}^{(2)}P_{\lambda}^{(2)}(cos\theta)}
            {\sum _{\lambda = 2, 4}A_{\lambda \lambda} P_{\lambda}(cos\theta)} 
                                                                      \label{Eqn_10}
\end{equation}

Above, A$_{\lambda \lambda}$ (A$_{22}$ and A$_{44}$) represents a$_2$  and a$_4$ coefficients, respectively. Legendre polynomials are denoted $P_{\lambda}(cos\theta)$ whereas $P_{\lambda}^{(2)}(cos\theta)$ are the associated Legendre polynomials. Coefficients $a_{\lambda}^{(2)}$ depend on the mixing ratio ($\delta$) of the transition, the corresponding F$_k$ coefficients can be found in Ref.~\cite{morinaga} and $\kappa _{\nu}$ values are given in table~II(b) of Ref.~\cite{faghana}.

\begin{figure}[h!]
\begin{center}
\includegraphics[width=\columnwidth]{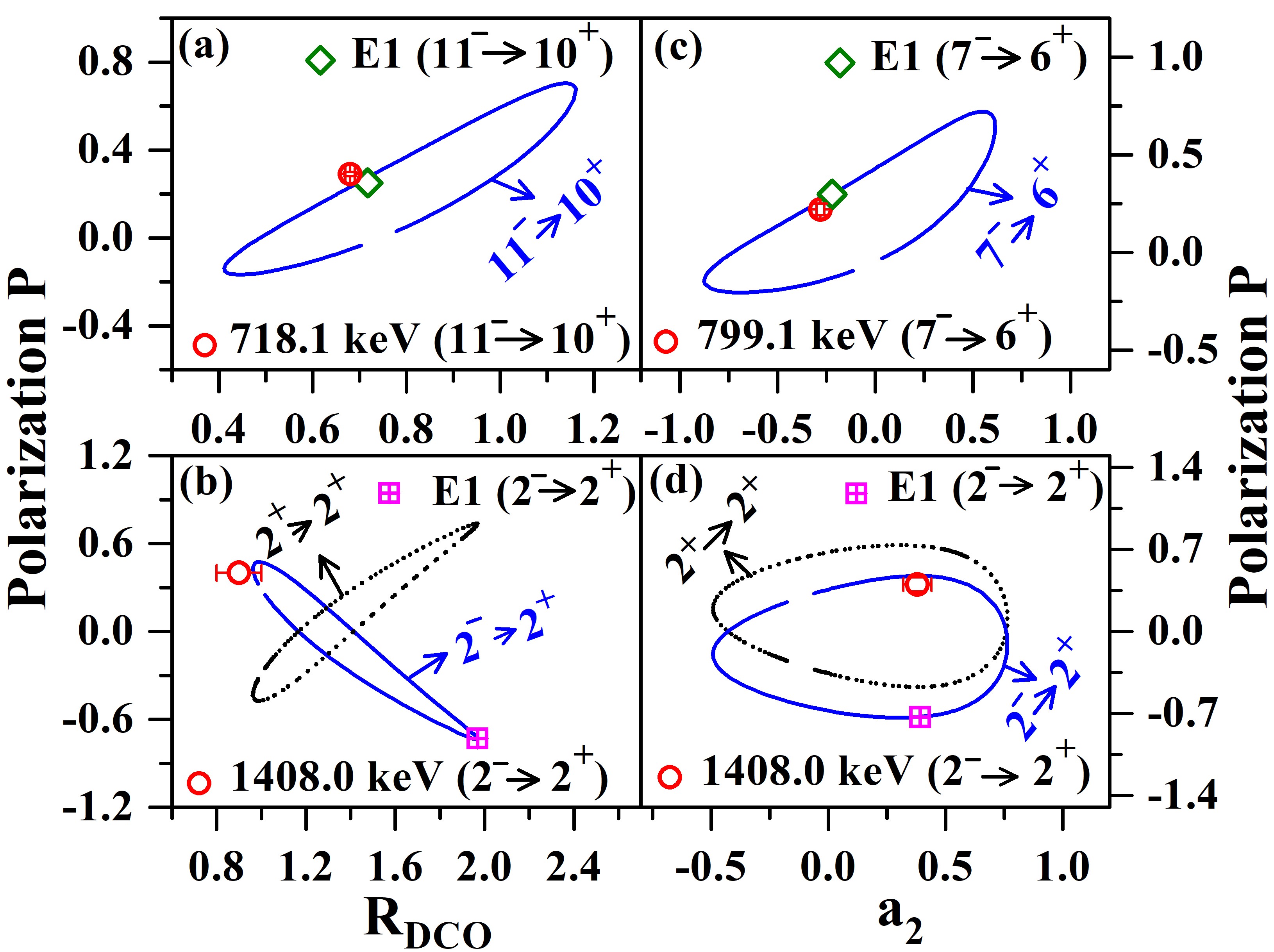}
\caption{{} The contour plots, made with calculated polarization, P against R$_{\rm DCO}$ or $a_2$ values are shown for the transitions in $^{152}$Sm for which the spin parity assignments are known. {The green diamonds represent the positions for streched E1 (I $\rightarrow$ I$^<$) transitions and pink squares represent the non-streched E1 (I $\rightarrow $ I)  transition. The red circles indicate the experimental data points. Panels (a) and (c) show contours for I $\rightarrow$ I$^<$ transitions whereas, panels (b) and (d) display contours for I $\rightarrow$ I transition. It is to be noted that the values of R$_{\rm DCO}$ for E1 transitions are different in the above two cases (I$\rightarrow$ I$^<$ and I $\rightarrow$ I)}. }
                                                                  \label{Fig_03}
\end{center}
\end{figure}

The plots of P$_{\rm calc}$ against calculated R$_{\rm DCO}$ or $a_2$ give rise to distinct contours corresponding to a transition of a given character (E$\lambda$ + M$\lambda'$) or (M$\lambda$ + E$\lambda'$) de-exciting a certain initial level with spin I$_i$ and parity 
$\pi_i$ to a certain final level with spin I$_f$ and parity $\pi_f$. The experimentally obtained P$_{\rm exp}$, R$_{\rm DCO}$ and $a_2$ values corresponding to the known and newly placed transitions in $^{152}$Sm were compared with these contours to determine the multipolarities of the transitions and to assign the spins and parities of the levels de-excited through these transitions.  Few such contours are shown in Fig.~\ref{Fig_03} for the transitions corresponding to known spins, multipolarities and mixing ratios.  Left(right) panels show P against  R$_{\rm DCO}$ or $a_2$. Fig~\ref{Fig_03}(a) shows the contour for the 718.1 keV ($11^-$ $\to$ $10^+$) E1-transition and Fig~\ref{Fig_03}(c) for the 799.1 keV ($7^-$ $\to$ $6^+$) E1-transition. The comparison of these values with the contours indicate their stretched E1-nature, as can also be found in  Ref.~\cite{ensdf}. 

Both P-vs.-R$_{\rm DCO}$ and P-vs-a$_2$ analyses have been performed for 1408.0\,keV transition  which is an I$\rightarrow$I transition. The results are shown in Fig.~\ref{Fig_03}, panels (b) and (d), respectively. The experimental data for this transition confirm its $2^-\rightarrow 2^+$~(E1+M2) nature. However, $\delta$ for this transition comes out to be higher than the known value in  Ref.~\cite{ensdf}. {The value of $\delta$ comes about +2.2 when calculated from the contour drawn with $\sigma /I$ = 0.33. It is observed that, in order to explain the experimental data on R$_{\rm DCO}$ and P for 1408.0~keV transition, the $\sigma$/I cannot exceed 0.4 for I = 2. The mixing ratio for this $\gamma$ ray comes down to +2.0 with a higher $\sigma /I$ = 0.4 considering the low spin (2$^-$) of the initial level (1529.80 keV,  Ref.~\cite{ensdf}).  In the present work, interpretation on the high mixing ratio for 1408.0~keV transition has not been attempted as it is not a part of the T$_d(2)$ band. However, it has been verified that the mixing ratio for this transition is determined} from conversion electron spectroscopy data. It is observed that there exists a wide variation in the electron intensity data for 1408.0~keV, measured by different authors, Ref.~\cite{goswami}. 

Similar contours of $a_4$-vs.-$a_2$ were also used for determination of the $\gamma$-ray multipolarity and the spin of the initial level except for parity.  In conclusion, the analysis involving known transitions validates the technique utilized in the present work for the determination of spin parity of the $\rm T_d(2)$-band. 

\noindent
Excluding the I$\rightarrow $I cases, stretched E2 transition satisfies the relations $a_2 > 0$ and $a_4 < 0$ and P positive. Similarly, a stretched E1 has $a_2 < 0$ and $a_4 = 0$ and P positive. For mixed transitions, the sign of P is guided by the level of mixing of higher multipoles and its value strongly depends on the $a_2$ and $a_4$ coefficients.

In case of I$\rightarrow $I transitions, $a_2>0$ and P negative indicates pure E1 transitions and $a_2>0$ and P positive indicates pure M1 transitions.  The values of polarizations are highly dependent on $\delta$ also in case of mixed I$\rightarrow $I transitions. The contours with I$\rightarrow $I transitions are very different from the contours with transitions I$\rightarrow$ I$^<$ (I$^< < $I) and with transitions I$\rightarrow $ I$^>$ (I$^> > $I) having same multipolarity. { It is also understood from the analysis of the $2^- \rightarrow 2^+$ 1408~keV transition. To be specific, for a $2^- \rightarrow 2^+$ E1 transition without any mixing, the R$_{\rm DCO}$ value is 1.98 in a $2^+ \rightarrow 0^+$ E2 gate, as calculated by ANCORR ($\sigma$/I = 0.33) and shown in Fig.~\ref{Fig_03}. This value is different from the one for an I$\rightarrow$ I$^<$ stretched E1 transition like 718.1~keV or 799.1~keV in E2 gate. Recall that the R$_{\rm DCO}$ and P values depend on I, $\sigma /I$ and the angular positions of the detectors.}

\section{Results and level scheme of the Candidate \boldmath $\rm T_d(2)$ band}
\label{ana}

%{\color{blue}{\bf(There are a large number of grammatical errors in the experimental section......Section is re-written)}}

\begin{figure}[h!]
\begin{center}
\includegraphics[width=\columnwidth]{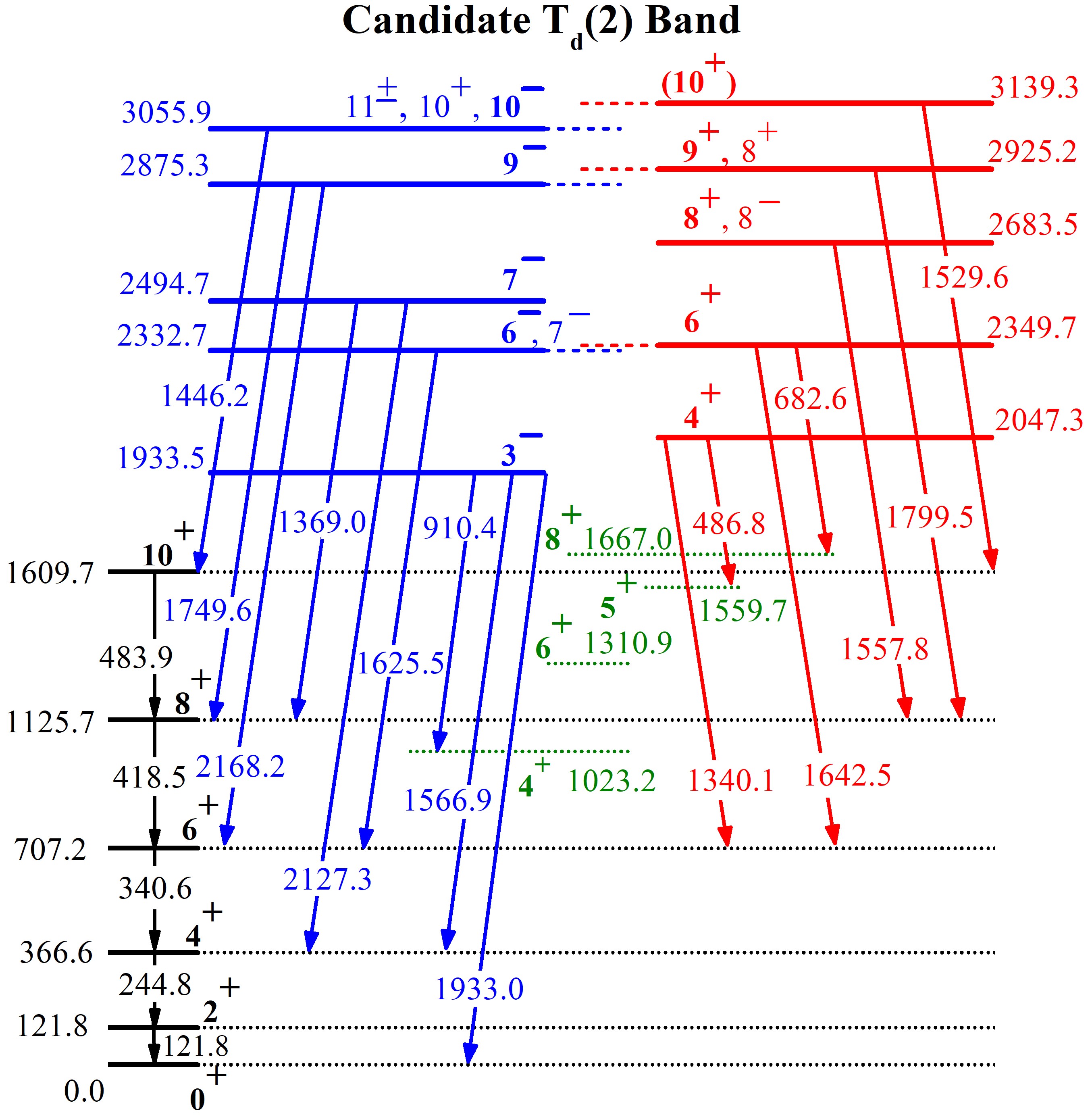}
\caption{Spin-parity structure of the T$_{\rm d}(2)$ band in $^{152}$Sm in full agreement with Eq.~(\ref{Eqn_06}). The negative parity branch is placed on the left-, and the positive parity one on the right-hand sides. The preferred spin parity assignments are shown in boldface. The preferences for the 2683.5~keV and 2925.2~keV levels are based on the proximity of experimental data points to a particular contour. The preferences for the 2332.7~keV, 3055.9~keV and 3139.3~keV levels are based on their agreement with the deduced quadratic energy-vs.-angular momentum dependence as presented in Fig.~\ref{Fig_18}, and may be seen as model dependent. All other level spin-parity assignments are the confirmed ones. For further specific details, see Sections IV A and IV B. Absence of intra-band transitions originating from T$_d(2)$ and transitions feeding any of the illustrated levels is worth noticing. The absence of inter-band transitions confirms strong structural differences between the tetrahedral and the other states. Let us emphasize the presence of parity doublets (dashed extentions).  }
                                                                  \label{Fig_04}
\end{center}
\end{figure}

\begin{figure*}[ht!]
\begin{center}
\includegraphics[width=\textwidth]{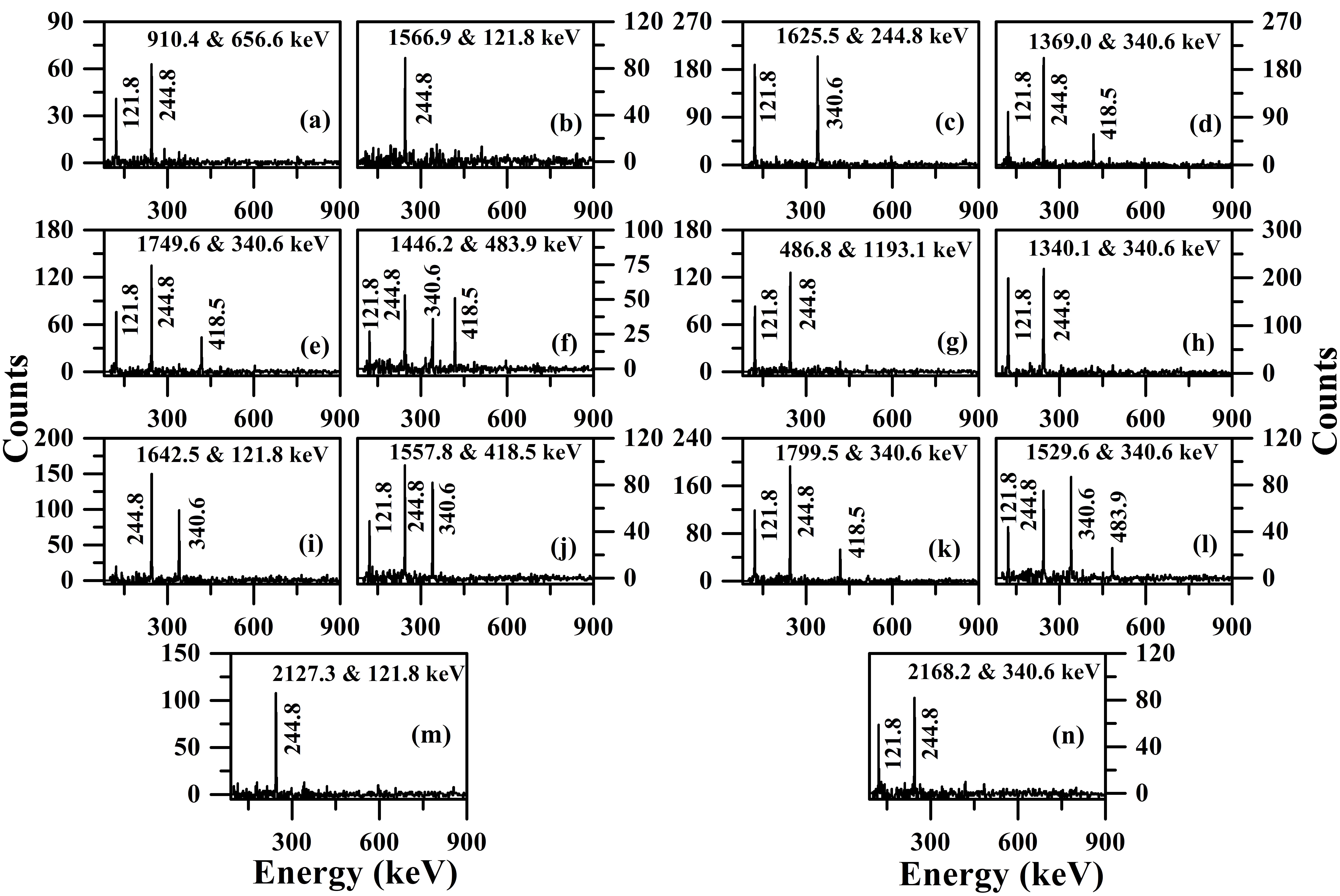}
\caption{Projections from triple $\gamma$ cube illustrating the analysis and subsequent observation of the coincidence relationships among the transitions placed in the level scheme of the $\rm T_d(2)$-band, shown in Fig.~\ref{Fig_04}, with the known $\gamma$-rays in $^{152}$Sm. See text for details.}
                                                              \label{Fig_05}
\end{center}
\end{figure*}

Figure~\ref{Fig_04} shows a sequence of levels that has been identified in $^{152}$Sm in the present work. The spin sequences shown in the Figure, referred to as Candidate $T_{\rm d}(2)$-Band satisfy  tetrahedral symmetry criteria in Eq.~(\ref{Eqn_06}) to a good approximation. 
The excitation energies of the levels (level energies), shown in the figure, are determined using energies of the decaying transitions, measured in the present work. The level energies in the $T_{\rm d}(2)$-band are listed in Table~\ref{tab1} along with their uncertainties. The measured characteristics of the $\gamma$-transitions depopulating this band are also shown in this table. Existence of these transitions was confirmed  on the basis of the results of coincidence analysis displayed in Fig.~\ref{Fig_05}. 

The $\rm T_{\rm d}(2)$-band low-lying levels decay to the ground-state~band or to a few other low-lying excited levels in $^{152}$Sm belonging to $\beta$ and $\gamma$ vibrational bands. However, no transitions were found to feed them, signifying differences in terms of the intrinsic structures of $\rm T_{\rm d}(2)$-band states compared to surrounding excited states, what can be seen as favoring tetrahedral symmetry interpretation and its very exotic structure with a few similarities (if any) with respect to the neighboring states.

The intensities I$_{\gamma}$ of the transitions in the $\rm T_{\rm d}(2)$-band, relative to the $4^+_1 \rightarrow 2^+_1$, 244.8\,keV transition (considered as $100\%$),  were determined from the total and gated projections of $\gamma$-$\gamma$ matrix and are shown in Table~\ref{tab1}. For 1933~keV transition, the intensity was obtained from singles (90$^{\circ}$) data.  No intra-band $E2$-transitions were found in the T$_{\rm d}(2)$-band within our observational limit and this supports the tetrahedral symmetry interpretation. The upper limit of intensities, for such missing intra-band $E2$-transitions, is lower than $0.010(3)\%$ compared to the intensity of the 244.8\,keV transition.

It turns out that, for most of the newly proposed levels in T$_{\rm d}(2)$-band, only I $ \rightarrow $ I and I $ \rightarrow$ I$^>$ transitions were observed. Transitions to lower spin levels (I $ \rightarrow$ I$^<$) were sought but could not be found probably due to their intensities below the detection limit.

In $^{152}$Sm, mostly for the low spin levels, higher intensities for the transitions of the type I $ \to$ I $^>$  are seen compared to those of I $ \to $ I$^<$ transitions. From  Ref.~\cite{ensdf} we deduce that 1579.42 keV $3^-$ level decays via 1212.94 keV  E1-transition, $3^- \to 4^+$ with I$^{\rm level}_\gamma$ = 100\%, compared to its decay via 1457.64 keV $(3^- \to 2^+)$ E1-transition with I$^{\rm level}_\gamma$ = 35.13\%, where I$^{\rm level}_\gamma$ is the branching intensity from a particular level, following the convention of Ref.~\cite{ensdf}. Similarly, transitions: 1730.2 keV, $3^-$ to a $4^+$ state, of 1764.32 keV, $5^-$ to a $6^+$ state, and of 1803.94 keV, $5^-$ to $6^+$ state, taken from   Ref.~\cite{ensdf}, proceed via the strongest decay branch to higher spin levels. State 2057.52 keV, $7^-$, also decays to $8^+$ state via a transition of I$^{\rm level}_\gamma$ = 65\%. The observed decays 
I$ \to $ I$^>$  from states of the T$_{\rm d}$(2)-band to the known $^{152}$Sm levels are therefore proceeding similarly. 

{  The intriguing issue of the observed dominance of the I $ \to$ I$^>$ type of transitions will be further addressed in Section V.D.}

\begin{figure}[h!]
\begin{center}
\includegraphics[width=\columnwidth]{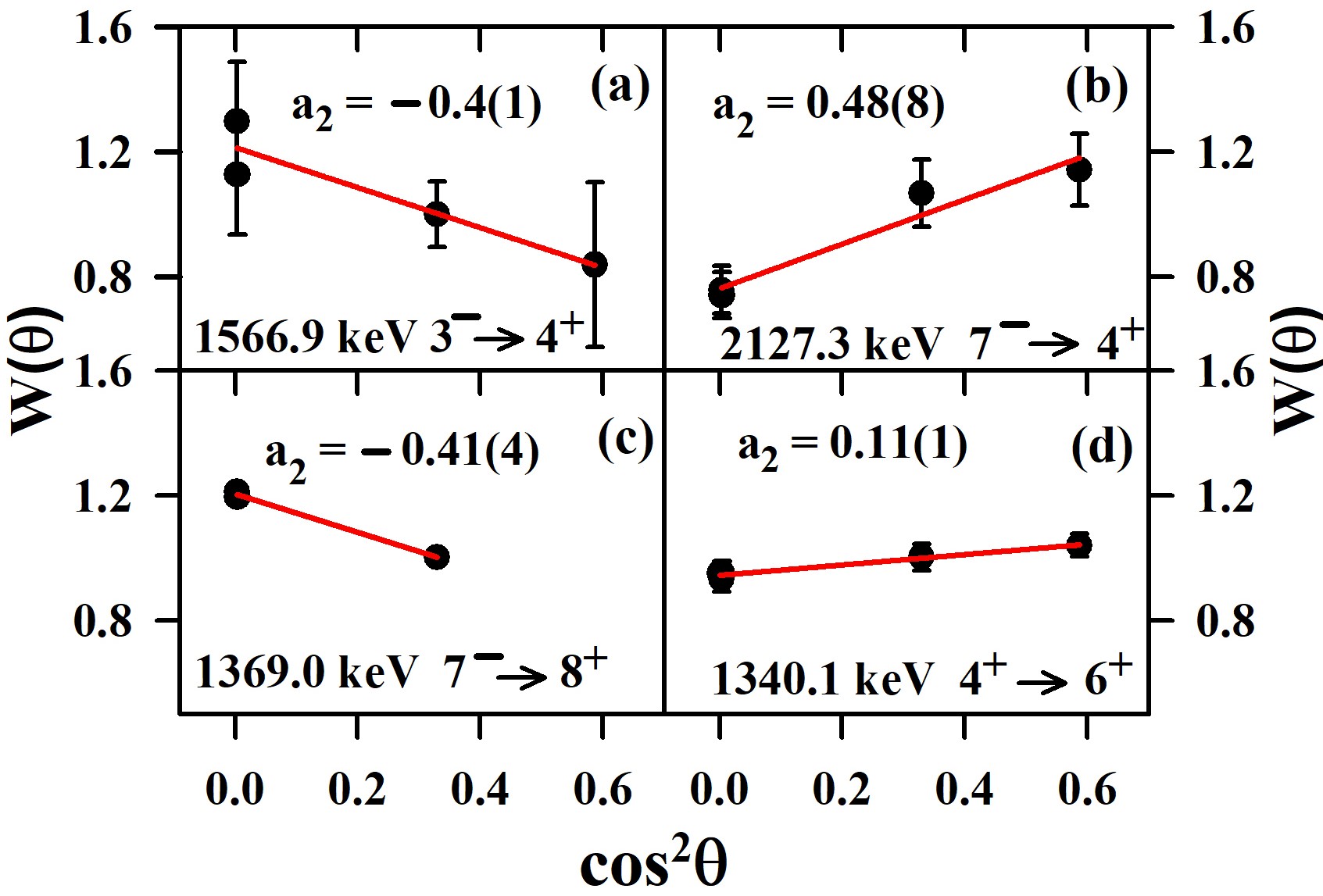}
\caption{The angular distributions W($\theta$) are  plotted as functions of $cos^2\theta$ for transitions E1, panels (a) and (c)), transitions E3, panel (b) and transitions E2, panel (d). The linear fit has been performed to determine the $a_2$ coefficients, after Ref.~\cite{e3}, where $a_2$ coefficients were determined for E3 transitions.} 
                                                                  \label{Fig_06}
\end{center}
\end{figure}

\begin{figure}[h!]
\begin{center}
\includegraphics[width=\columnwidth]{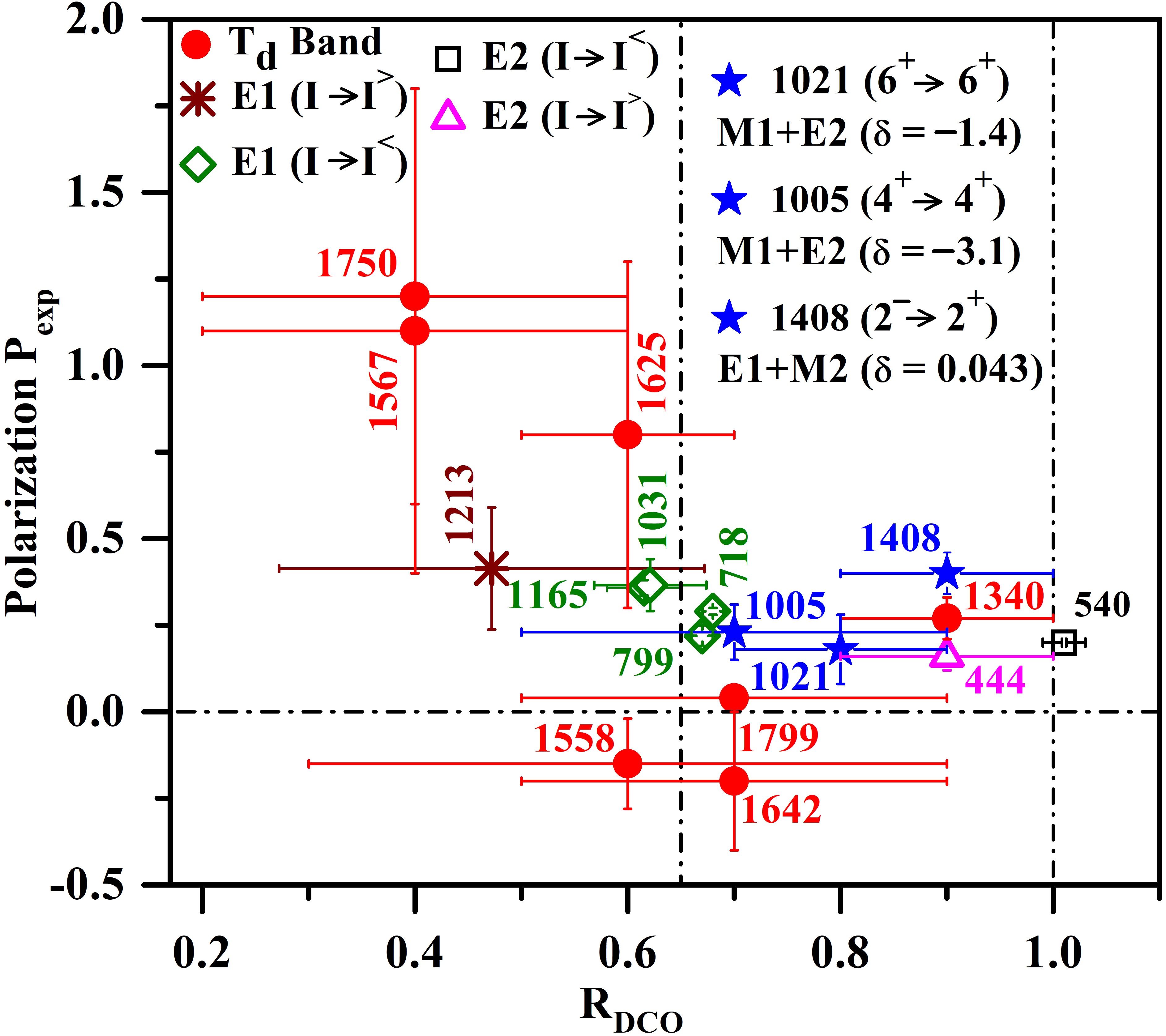}
\caption{The measured R$_{\rm DCO}$ and polarization P$_{exp}$ values are shown for the known transitions in $^{152}$Sm and the new ones placed in the T$_{\rm d}(2)$-band. The vertical lines are drawn to guide the eye for pure dipole and quadrupole transitions. The dashed horizontal line indicates P$_{exp}$ = 0.}
                                                                  \label{Fig_07}
\end{center}
\end{figure}

\begin{figure}[h!]
\begin{center}
\includegraphics[width=\columnwidth]{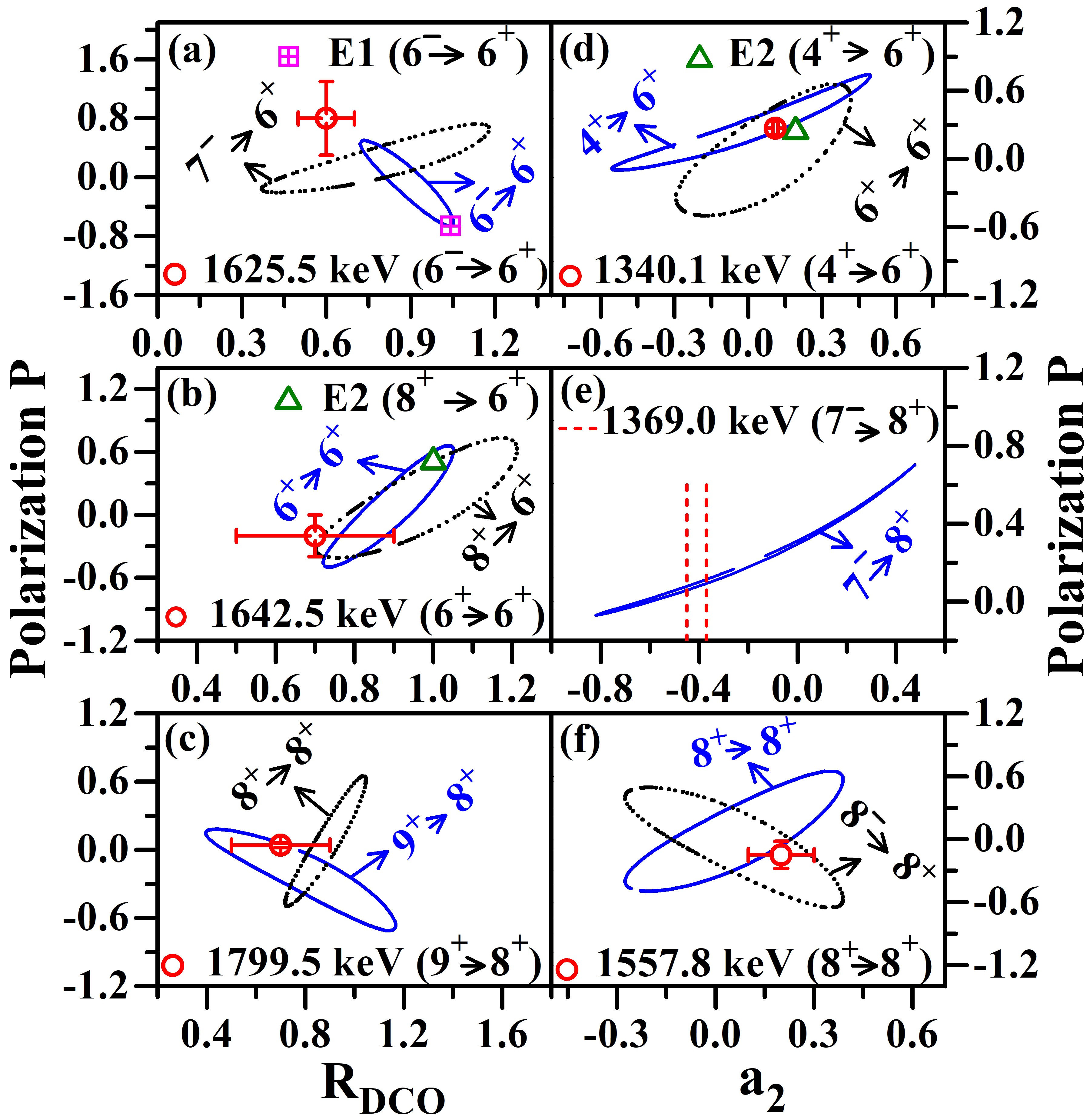}
\caption{ The contour plots made with calculated R$_{\rm DCO}$/$a_2$ and Polarization are shown for the transitions of the T$_{\rm d}(2)$-band. { The green triangles represent the stretched E2 transitions and pink square the non stretched E1 (I $\rightarrow$ I) ones. The red circles denote the experimental data points.  All the panels except panel (e) contain contours for the I $\rightarrow$ I transitions. See text and Fig.~\ref{Fig_03} for details on I $\rightarrow$ I transitions.}}
                                                                 \label{Fig_08}
\end{center}
\end{figure}

The spins and parities of the levels of $\rm T_d(2)$-band measured in the present work are collected in Table~\ref{tab1} following the results of analysis displayed in Figs.~\ref{Fig_06}, \ref{Fig_07} and \ref{Fig_08}. For some cases, where definite spin parity assignments were not feasible, all the possible spin parity values are indicated.


The following text describes all the levels of the $\rm T_d(2)$-band.   

%%%%%%%%%%%%%%%%%%%%%%%%%%%%%%%%%%%%%%%%%%%%%%%%%%%%%%%%%%%%%%%%%%%%%%%%%%%%%%%%

\subsection{Negative Parity Levels of the \boldmath Candidate T$_{\rm d}$(2) Band}

\subsubsection{1933.5~keV, $3^-$ level}

This level was already known in as 1933.30(5)~keV,  Ref.~\cite{ensdf}. In the present work, its excitation in $^{152}$Sm was confirmed and placed in the  T$_{\rm d}(2)$-band following the observation of 910.4(1)~keV, 1566.9(2)~keV and 1933.0(8)~keV transitions. The coincidence relations for the former two transitions, which were known in  Ref.~\cite{ensdf} as 910.38(7)\,keV and 1566.82(8)\,keV, are shown in Figs.~\ref{Fig_05}(a) and (b). The placement of 1933.0(8)~keV transition was based on the analysis of singles data and comparison to the $\gamma-\gamma$ matrix.

\begin{figure*}[ht]
\begin{center}
\includegraphics[width=\textwidth]{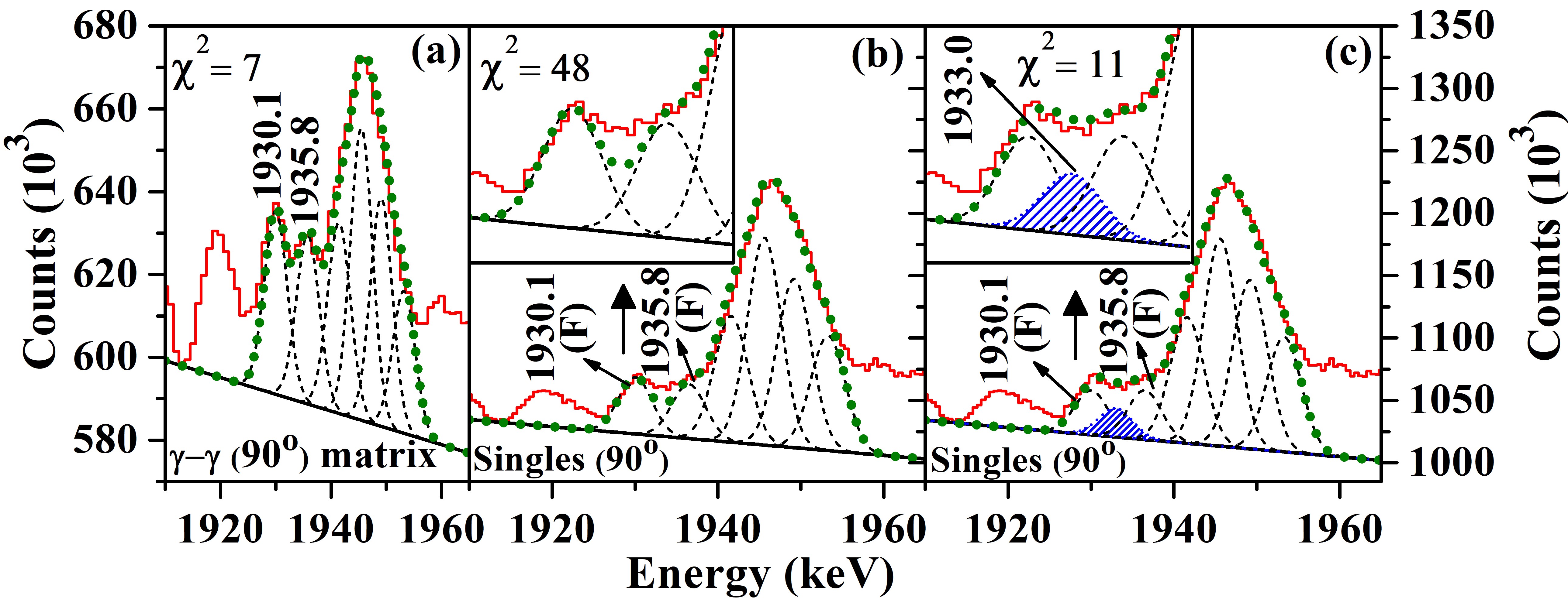}
\caption{Selected projections from $\gamma$-$\gamma (90^{\circ})$ matrix, (a), and  singles (90$^{\circ}$) (b) and (c),  are shown. In panel (a), the 90$^{\circ}$  projection from $\gamma$-$\gamma (90^{\circ})$ matrices (red solid line) is fitted (green dotted line) to determine the energies of the observed peaks (black dashed lines) around 1933\,keV. In panels (b) and (c), the singles (90$^{\circ}$) (red solid line) spectra are shown and peaks are fitted (green dotted lines) fixing the peaks observed through fitting in $\gamma$-$\gamma(90^{\circ})$ matrix data of panel (a). The zoomed views in panel (b) and (c) are displayed in the insets to show that the singles (90$^{\circ}$) projection can be best fitted considering the presence of a 1933.0\,keV peak (shown with blue hashed region). The $\chi^2$ values are also given.}
                                                              \label{Fig_09}
\end{center}
\end{figure*}

No 1933.5~keV transition energy was found in the 90$^{\circ}$ projection from the $\gamma$-$\gamma(90^{\circ})$ matrix data. In the region around 1933.5~keV, two peaks, {\em viz.}~1930.1(1)~keV and 1935.8(2)~keV were found in this projection, as shown in Fig.~\ref{Fig_09}(a). One of these transitions (1930.1\,keV, cited as 1930.05(7)\,keV in Ref.~\cite{ensdf} was known earlier and from the present data one can confirm that  the 1930.1~keV and 1935.8~keV transitions feed the 121.8~keV and 1125.7~keV levels in $^{152}$Sm, respectively.

Analyzing  the singles (90$^{\circ}$) data we deduced the existence of an intermediate transition with an energy between 1930.1~keV and 1935.8~keV. In order to find the energy of this transition,  least squares fitting to the singles (90$^{\circ}$) spectrum was performed with the results shown in Figs.~\ref{Fig_09}(b) and \ref{Fig_09}(c). This was done without and with the presence of a transition between 1930.1~keV and 1935.8~keV to verify the quality of least squares fitting. For this purpose, global background and multiple Gaussian peaks with expected FWHM at around 2\,MeV were used. The peak energies were fixed at values observed in the $\gamma-\gamma(90^{\circ})$ data of Fig.~\ref{Fig_09}(a) with a $\chi^2$ value of 7, except the energy of the intermediate transition.

We find that the fitting without considering any peak between 1930.1~keV and 1935.8~keV transitions results in a $\chi^2$ value of 48 compared to 11 obtained when considering a peak between 1930.1~keV and 1935.8~keV. The energy of this new peak observed in the singles 90$^{\circ}$-spectrum was found to be 1933.0(8)~keV. Observation of this transition only in the singles from 90$^{\circ}$ data suggests its origin at the 1933.5\,keV level  of T$_{\rm d}(2)$-band, feeding the ground state.

\begin{figure}[h!]
\begin{center}
\includegraphics[width=\columnwidth]{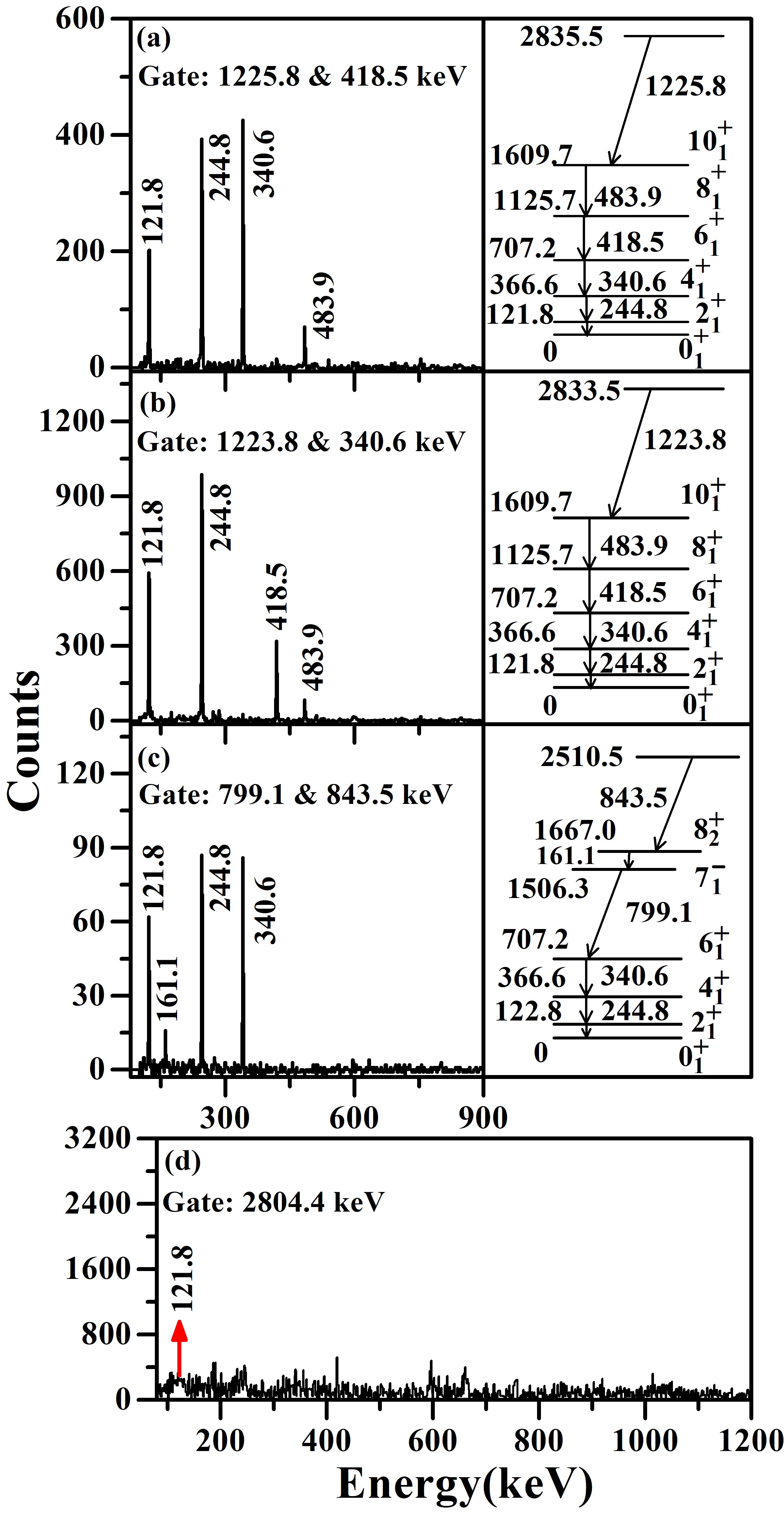}
\caption{Projections from triple $\gamma$ cube (a), (b) and (c) and $\gamma-\gamma$ matrix (d) are shown to illustrate the analysis results and observation of the coincidence relationships  among transitions depopulating the levels 1933.30\,keV, 2348.76\,keV and 2925.5\,keV, Ref.~\cite{ensdf}, but have been placed differently or not placed in the present work. Note that gates of 0.1 keV were applied to select the transitions of interest;   see text for details.}
                                                              \label{Fig_10}

\end{center}
\end{figure}

It is to be noted that,  the presence of a 1226.32(7) keV (Ref.~\cite{ensdf}) transition, earlier known as decaying from the 1933.5 keV level, could not be confirmed in the present work. A 1225.8(1)~keV transition was found in the present data, which shows coincidence with all of the transitions of the ground state band up to 10$^+$, 1609.7~keV level, Fig.~\ref{Fig_10}(a). This suggests that the 1225.8~keV transition does not depopulate the 1933.5~keV level. 

The spin-parity assignment for the 3$^-$ level of the T$_{\rm d}(2)$-band originates from the E1 multipolarity of  the 1566.9\,keV transition. This is based on $a_2 < 0$, Fig.~\ref{Fig_06}(a), and R$_{\rm DCO}$ value close to that of 1212.94\,keV ($3^-\rightarrow~4^+$) E1-transition and distinctly positive P value, Fig.~\ref{Fig_07}. As discussed earlier, the known non-yrast 3$^-$ and 5$^-$ levels in $^{152}$Sm decay to the ground-state band, having stronger transitions to the higher spin states.  Therefore, the $I^{\pi}$ of 1933.5\,keV level is rather safely assigned as 3$^-$ (rather than 5$^-$). This is also confirmed by the fact  that the 1226.32\,keV transition (earlier proposed as depopulating 1933.5 keV level to the 6$^+$, 707.2\,keV level)  does not originate from the 1933.5\,keV level and was attached to 1609.7 keV, 10$^+$ level. The multipolarity of 1933.0(8)~keV is indirectly assigned as E3. 

\subsubsection{2332.7~keV $(6^-,7^-)$ level}

This level was found in the present work on the basis of the coincidence relationships followed by the newly observed 1625.5(1)~keV transition, shown in Fig.~\ref{Fig_05}(c). The spin parity assignment of this excitation is solely dependent on the R$_{\rm DCO}$ and P obtained for 1625.5~keV transition. For this transition, P is distinctly positive and gives support for an 6$^- \rightarrow 6^+$ (E1+M2)-assignment cf.~Fig.~\ref{Fig_08}(a). The discussed transition, however, also shows a possibility for having 7$^-$ $\rightarrow$ 6$^+$ assignment, which could not be eliminated due to large error bars in the experimental results. The  assignment in the form of an alternative $(6^-,7^-)$ is retained for the 2332.7~keV level. 

\subsubsection{2494.7~keV, $7^-$  level}

The 2494.7~keV, $7^-$ level was placed in the $\rm T_d(2)$-band in the present work. The 1369.0(1)~keV and 2127.3(4)~keV transitions were found to depopulate this level to the 8$^+$ and 4$^+$ levels of the ground state band, respectively. The coincidence of these transitions is illustrated in Figs.~\ref{Fig_05}(d) and (m).  

The W($\theta$) analysis for 2127.3~keV transition was performed knowing that the slope of $W(\theta)$-vs.-$cos^2(\theta)$ relation is higher for E3 transitions compared to E2 transitions,~Ref.~\cite{e3}, resulting from their high positive $a_2$ values. The $E3$ assignment for 2127.3\,keV transition follows from the angular distribution measurement shown in Fig.~\ref{Fig_06}(b), giving $a_2$ = 0.48(8) compared to 0.11(1) for 1340.1~keV E2-transition (discussed later). The $a_2$ value for the pure E3 ($7^- \rightarrow 4^+$) decay (2127.3\,keV) is 0.55, as calculated by ANCORR. This confirms the 7$^-$ assignment for the 2494.7\,keV level. The multipolarity of 1369.0\,keV transition was identified as E1+M2, given its $a_2$ value, Fig.~\ref{Fig_06}(c), which falls within the contour of 7$^- \rightarrow 8^+$ decay, Fig.~\ref{Fig_08}(e), although the P$_{exp}$ for this transition could not be determined.

\subsubsection{2875.3~keV, $9^-$ level}

Placement of 2875.3~keV, $9^-$ level in T$_d(2)$-band is based on the observation of the 1749.6(1)~keV and 2168.2(5)~keV transitions, whose coincidence relations are displayed in Figs.~\ref{Fig_05}(e) and 5(n), respectively. The 1749.6\,keV transition was assigned as E1, based on its R$_{\rm DCO}$, $a_2 < 0$ and $P > 0$ as can directly be seen from Fig.~\ref{Fig_07} and Table~\ref{tab1}. This gives rise to 9$^-$ assignment for the 2875.3\,keV level of the T$_d(2)$-band. Consequently, the E3-nature of the 2168.2~keV transition can be considered deduced.

\subsubsection{3055.9~keV, $10^{(\pm)}$ or $ 11^{(\pm)}$ level }

We observed a 1446.2(1)~keV transition showing coincidence with the transitions of ground-state band. They depopulate the 1609.7~keV level, as it is visible from Fig.~\ref{Fig_05}(f).  The 3055.9~keV level of the T$_{\rm d}(2)$-band was placed on the basis of this observation. The R$_{\rm DCO}$ value for this transition indicates its dipole character (I $\rightarrow I^<$) or mixed nature (I $\rightarrow I$). However, no polarization measurement could be performed due to weak statistics. Accordingly, spin parity assignment for this level could not be completed and the following assignments are considered possible: $I^\pi=10^+,10^-,11^+$ and $11^-$.

\subsection{Positive Parity Levels of \boldmath the  T$_d(2)$-Band}

\subsubsection{2047.3\,keV, $4^+$ level}

This level was found from the coincidence relations between the 486.8(3)~keV and  the 1340.1(1)\,keV transitions shown in Figs.~\ref{Fig_05}(g) and 5(h). A 2046(10)\,keV level is known from  Ref.~\cite{ensdf}, however without definite I$^{\pi}$ assignment. The de-exciting $\gamma$ transitions from that level were known as 1339.33(11)~keV and 486.2(2)~keV according to Ref.~\cite{ensdf}. The energies of these  transitions are within 1~keV when compared with the energies of transitions measured in the present work. Accordingly, the  2047.3\,keV, $4^+$ level is considered the same as the one adopted earlier by  Ref.~\cite{ensdf}. 

The R$_{\rm DCO}$ and P values for the 1340.1\,keV transition are similar to the 444\,keV, ($2^+\rightarrow 4^+$) E2 transition, Fig.~\ref{Fig_07}, and its $a_2$ value is positive as seen from Fig.~\ref{Fig_06}(d). This confirms the I$^\pi =4^+$ assignment for the 2047.3\,keV level. This is also confirmed from the P vs. $a_2$ contour shown in Fig.~\ref{Fig_08}(d) as the experimental data point coincides with calculated point for stretched E2.

\subsubsection{2349.7~keV, $6^+$ level}

This $6^+$ level was placed in the $\rm T_d(2)$-band following the observation of the 682.6(1)~keV and 1642.5(2)~keV transitions in coincidence with the known $\gamma$ transitions of $^{152}$Sm. The coincidence relation of  the 1642.5~keV $\gamma$-ray is displayed in Fig.~\ref{Fig_05}(i).

The 2348.76(7)\,keV level is known in  Ref.~\cite{ensdf}. However, the 843.36(17)\,keV and 1223.47(9)\,keV transitions, earlier known according to Ref.~\cite{ensdf}, could not be confirmed in our work, as discussed below.

We found two transitions around 1223.47\,keV  (Ref.~\cite{ensdf}) transition: 1223.1(1)~keV and 1223.8(1)~keV. 
Gated projections were studied employing RADWARE with gate width of 0.1~keV to reduce the contamination from the close lying $\gamma$-peaks. The lower one, 1223.1(1)~keV was already known in  Ref.~\cite{ensdf} as 1223.16(9)\,keV depopulating the 1929.93(6)\,keV (Ref.~\cite{ensdf}) level. 
This interpretation was also confirmed in the present work and therefore, the discussed transition cannot be the one depopulating the 2349.7\,keV level. The higher one, 1223.8(1)~keV transition, shows coincidence relationships in favor of its connection to 1609.7~keV level of the ground-state band, see Fig.~\ref{Fig_10}(b). Therefore, we conclude that, no 1223.47\,keV  transition was identified as depopulating the 2349.7\,keV level of the T$_{\rm d}(2)$-band. 

%\st{attribution to} 

The 843.36(17)\,keV transition known earlier was found as 843.5(2)~keV in the present work,  depopulating the 2510.5~keV level and feeding the 1667.0~keV level. This latter one subsequently decays via the 161.1~keV transition (160.8~keV, Ref.~\cite{ensdf}). The placement of  the 843.5\,keV transition in the level scheme of $^{152}$Sm was confirmed from the triple coincidence displayed in Fig.~\ref{Fig_10}(c) and the relative intensities of 161.1~keV and 843.5\,keV transitions.

The (2349.7~keV) level is known in  Ref.~\cite{ensdf}, however without a spin-parity assignment.  The R$_{\rm DCO}$ and P values for the 1642.5\,keV transition, Fig.~\ref{Fig_07}, suggests $6^+\rightarrow 6^+$ (M1+E2)-assignment  as also reflected in the 
P-vs.-R$_{\rm DCO}$ contours shown in Fig.~\ref{Fig_08}(b). The experimental data point for the 1642.5\,keV transition with its error bar also falls on the contour of $8^+\rightarrow 6^+$ (E2+M3). However, this latter assignment suggests a very high mixing of $\delta = 0.9$ M3-transition, which is very unlikely in general. In addition, the E2 nature of  the 682.6\,keV transition also follows from its DCO ratio of 0.9(1) which is similar to the one calculated from ANCORR and equal 0.95. This supports the $I^{\pi}=6^+$ assignment in contradiction to the $I^{\pi}=7^+$ or 8$^+$ attribution to the 2349.7 keV level.

\subsubsection{2683.5~keV, $8^{(+)}$ level}

\begin{figure}[ht!]
\begin{center}
\includegraphics[width=\columnwidth]{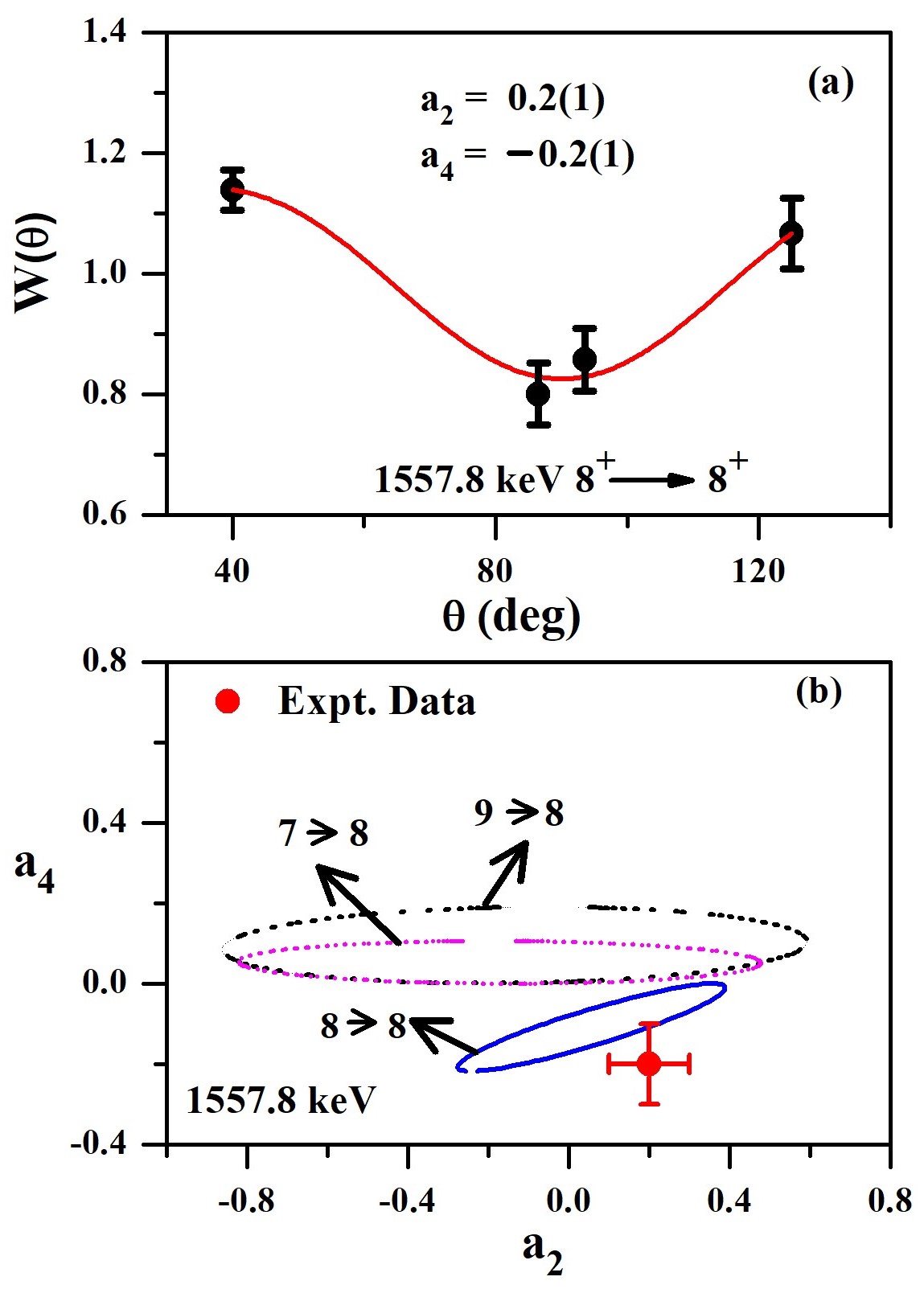}
\caption{The angular distribution fits for an $I\rightarrow I$, M1+E2 transition (1557.8\,keV) (a) observed in the T$_{\rm d}(2)$-band. Panel (b) shows the $a_2$-vs.-$a_4$ contour, used in the 8$^+$ spin assignment of the 2683.5\,keV level, where the 10$\to$8 contour (E2 transition)  is not shown because 1557.8\,keV has negative P value.  }
                                                                  \label{Fig_11}
\end{center}
\end{figure}

Figure~\ref{Fig_05}(j) shows the coincidence relations of the 1557.8(1)~keV transition  from the 2683.5~keV, $8^+$ level. The angular distribution analysis for this transition is shown in Fig.~\ref{Fig_11}(a) and the $a_2$-vs.-$a_4$ contours, shown in Fig.~\ref{Fig_11}(b)  confirm the I = 8 assignment for this level.   
However, the P-vs.-$a_2$ contours of Fig.~\ref{Fig_08}(f)  show that the measured value, although  falling on the $8^+\rightarrow 8^+$ contour, may also support $8^-\rightarrow 8^+$ assignment due to large experimental error bar in the $a_2$ value. 

\subsubsection{2925.2~keV, $(9)^{+}$ or 
 $(8)^{+}$  level}

The 1799.5(1)\,keV transition, found in the present work, decays from the 2925.2~keV level following the cube gates projected in Fig.~\ref{Fig_05}(k).

The 2925.2\,keV level of the $\rm T_d(2)$-band is suggested to differ from the 2925.5(10)\,keV level, which was known in  Ref.~\cite{ensdf} from $\beta^-$ decay, cf.~Ref.~\cite{mach}. This is confirmed  by examining the gated projection of  the 2803.7(10)\,keV (Ref.~\cite{ensdf}) transition which was known as the only transition depopulating the 2925.5\,keV level  to the 121.8\,keV level. It is observed that the transition 2803.7\,keV (Ref.~\cite{ensdf}) corresponding to 2804.4(8)\,keV,  in the present work, has no coincidence with the 121.8\,keV $\gamma$-ray, see Fig.~\ref{Fig_10}(d).

The experimental R$_{\rm DCO}$ and P$\approx$0 for 1799.5(1)~keV transition indicates the $9^+\rightarrow 8^+$ (M1+E2)-character of this transition. However, the large error bar in the R$_{\rm DCO}$ value also touches the $8^+\rightarrow 8^+$ 
(M1+E2)-contour, as shown in Fig.~\ref{Fig_08}(c), resulting with uncertainties in the present identification.

\subsubsection{3139.3~keV, $(10^+)$ level}

The 3139.3(2)~keV level was found from the cube gates of 1529.6(1)~keV displayed in Fig.~\ref{Fig_05}(l).  For spin-parity assignments of this level, our experimental information is rather incomplete allowing only for qualitative argumentation. This level, like other levels placed in the $\rm T_d(2)$-band satisfies the expected parabolic relationship expected from Eq.~\ref{Eqn_06}, however its spin-parity identification remains uncertain; due to closeness of the position of this level to the expected position of the $10^+$ member of the doublet we believe that this is the more likely possibility -- without numerical arguments.\\[2mm]

The data presented so far allow the spin-parity sequences to be chosen that conform to an interpretation of the levels as having the characteristic for rotational bands generated in the presence of the tetrahedral symmetry, as given by Eq.~(\ref{Eqn_06}). Further physical consequences of these choices, shown in bold on Fig.~\ref{Fig_04}, will be discussed next.

\begin{table*}[h!]
\begin{center}
\caption{The level energies (E$_i$) of T$_d(2)$ band (Fig.~\ref{Fig_04}) are shown along with the details on the $\gamma$ transitions (E$_{\gamma}$) depopulating these levels. The energies of the $\gamma$ transitions and levels are shown with their uncertainties. The uncertainties in the level energies are evaluated using the method of addition using the uncertainties measured for the energies of the observed $\gamma$ rays. The intensities (I$_{\gamma}$), angular distribution coefficients (a$_2$, a$_4$), R$_{\rm DCO}$ and Polarization for the transitions are given. The listed intensities are relative to the $4^+_1 \rightarrow 2^+_1$, 244.8\,keV transition, considered as $100\%$. The second last column contains proposed multipolarities for a particular level. The confirmed/adopted spin parity values (I$_i^{\pi}$) for the levels of $\rm T_d(2)$ band are shown in third column and the other possibilities have also been indicated, if there is any, in the last column.
 See text for details.
}  
\begin{tabular}{cccccccccccccc}
\hline
\hline
$E_i$&$E_f$&$I_i^{\pi}$&$\rightarrow I_f^{\pi}$&$E_{\gamma}$ & $I_{\gamma}$&$a_2$&$a_4$&R$_{\rm DCO}$&$\Delta _{IPDCO}$&Polarization&Multipolarity&Other\\
&&&&&&&&&&P$_{expt.}$&(proposed)& possible\\
(keV)&(keV)&&&(keV)&&&&&&($\Delta _{IPDCO}$/Q)&&I$_i^{\pi}$\\
\hline
1933.5(8)&1023.22(4)& {3$^-$}&$\rightarrow 4^+$&910.4(1)&0.20(2)&&&&&&\\
&366.60(2)&&$\rightarrow 4^+$&1566.9(2)&0.18(2)&-0.4(1)&0.01(22)&0.4(2)&0.4(3)&1.1(7)&E1&\\
&0.0&&$\rightarrow 0^+$&1933.0(8)&0.05(1)&&&&&&E3\footnotemark&\\
%&&&&&&&&&&&&\\
2047.3(3)&1559.71(4)&{4$^+$}&$\rightarrow 5^+$&486.8(3)&0.71(13)&&&0.4(2)&&&(M1)&\\
&707.21(4)&&$\rightarrow 6^+$&1340.1(1)&0.77(6)&0.11(1)&-0.02(4)&0.9(1)&0.11(3)&0.27(6)&E2&\\
%&&&&&&&&&&&&\\
2332.7(2)&707.21(4)&{(6)$^{-}$}&$\rightarrow 6^+$&1625.5(1)&0.35(2)&&&0.6(1)&0.3(2)&0.8(5)&E1+M2&7$^-$\\
%&&&&&&&&&&&\\
2349.7(2)&1667.97(6)&6$^+$&$\rightarrow 8^+$&682.6(1)&0.14(2)&&&0.9(1)&&&E2&\\
&707.21(4)&&$\rightarrow 6^+$&1642.5(2)&0.12(1)&&&0.7(2)&-0.07(7)&-0.2(2)&M1+E2&\\
%&&&&&&&&&&&\\
2494.7(4)&1125.75(6)&{7$^-$}&$\rightarrow 8^+$&1369.0(1)&0.10(1)&-0.41(4)&&&&&E1+M2&\\
&366.60(2)&&$\rightarrow 4^+$&2127.3(4)&0.07(1)&0.48(8)&&&&&E3&\\
%&&&&&&&&&&\\
2683.5(2)&1125.75(6)&{8$^{(+)}$}&$\rightarrow 8^+$&1557.8(1)&0.46(2)&0.2(1)&-0.2(1)&0.6(3)&-0.06(5)&-0.15(13)&M1+E2&8$^-$\\
%&&&&&&&&&&\\
2875.3(5)&1125.75(6)&{9$^-$}&$\rightarrow 8^+$&1749.6(1)&0.15(1)&-0.33(6)&&0.4(1)&0.4(2)&1.2(6)&E1&\\
&707.21(4)&&$\rightarrow 6^+$&2168.2(5)&0.05(1)&&&&&&(E3)$^a$&\\
%&&&&&&&&&&\\
2925.2(2)&1125.75(6)&{(9)$^+$}&$\rightarrow 8^+$&1799.5(1)&0.27(2)&&&0.7(2)&0.02(1)&0.04(2)&M1+E2&8$^+$\\
%&&&&&&&&&&\\
3055.9(2)&1609.69(7)&{(10$^{-})$}&$\rightarrow 10^+$&1446.2(1)&0.14(2)&&&0.5(2)&&&(E1+M2)&10$^+$,11$^{\pm}$\\
%&&&&&&&&&&&&\\
%&&{\color{blue}}&&&&&&&&&&\\
%&&&&&&&&&&\\
3139.3(2)&1609.69(7)&{(10$^+$)}&$\rightarrow 10^+$&1529.6(1)&0.17(3)&&&&&&(M1+E2)&\\
\hline
\hline
\end{tabular}
\footnotetext{Assigned indirectly from $I^{\pi}$ assignment of initial level}
\label{tab1}
\end{center}
\end{table*}

\section{Quantum Symmetries Behind OBSERVED BANDS \boldmath T$_{\rm d}(1)$ and T$_{\rm d}(2)$ }
\label{theory}

Let us begin by illustrating the results of the total energy calculations which will be treated as the starting point of the interpretation and discussion of microscopic background of the underlying rotational exotic band properties.  To describe nuclear shapes which represent the two competing nuclear symmetries, T$_{\rm d}$ and O$_{\rm h}$ introduced earlier, we employ the first order tetrahedral, $t_1$, and octahedral, $o_1$, shape coordinates, cf.~Eqs.~(\ref{Eqn_02}-\ref{Eqn_05}). 

\begin{figure}[h!]
\begin{center}
\includegraphics[width=\columnwidth]{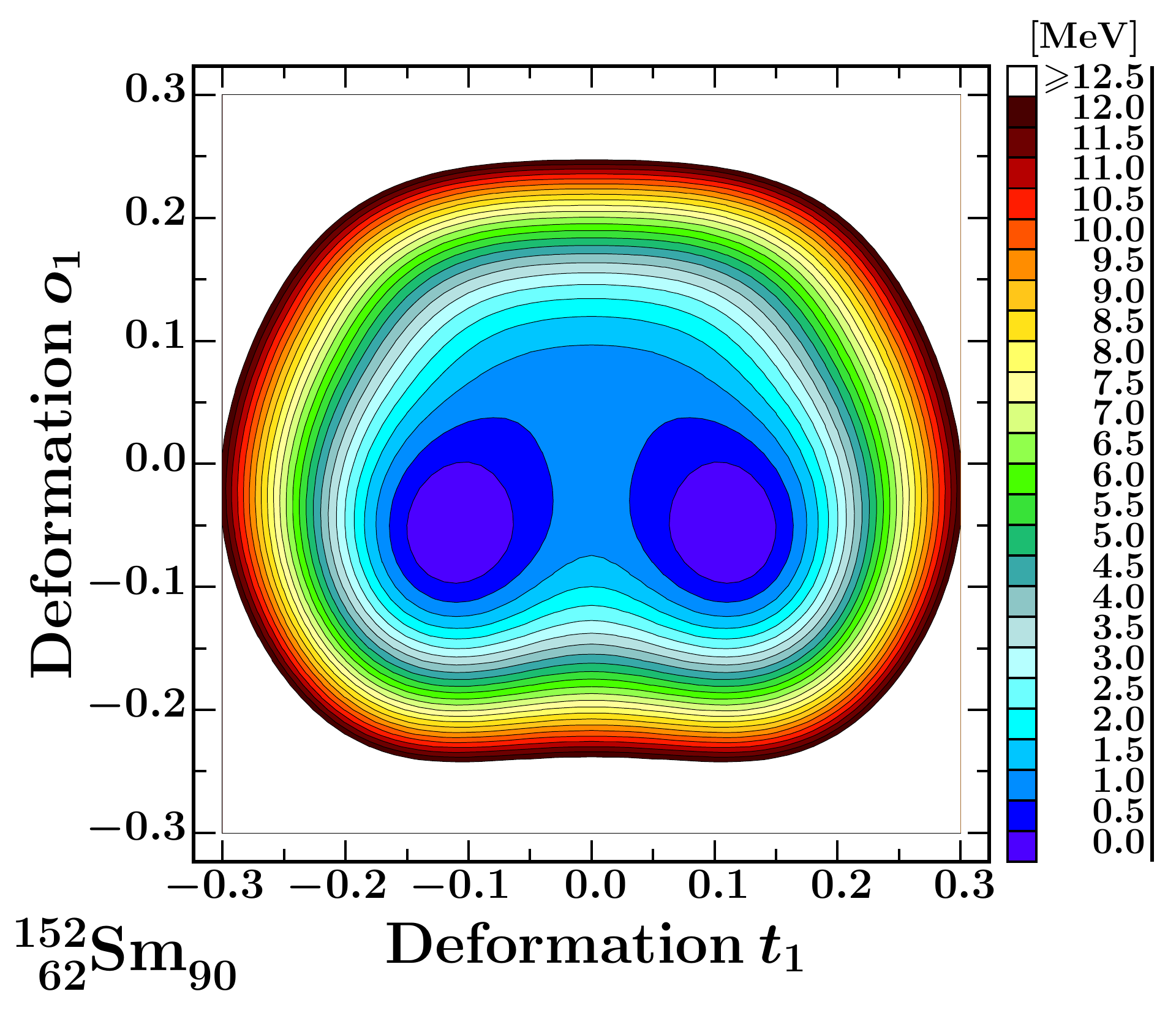}
\caption{Total energy from macroscopic-microscopic method after Ref.~\cite{GDD21}, as a function of tetrahedral ($t_1$) and octahedral ($o_1$) deformations, Eqs.~(\ref{Eqn_02},\ref{Eqn_04}), cf.~also~Ref.~\cite{JDEPJ} for mathematical details and discussion. Observe that static equilibrium energy minima correspond to non-vanishing tetrahedral and octahedral deformations ($t_1=\pm0.12$, $o_1=-0.06$) suggesting coexistence of both symmetries. 
}
                                                                      \label{Fig_12}
\end{center}
\end{figure}

Interested reader will find mathematical details in the recent review article, Ref.~\cite{JDEPJ} and in the references therein. Figure \ref{Fig_12} shows the resulting potential energy surface. The landscape indicates presence of twin minima at non-vanishing $t_1 = \pm 0.12$ and $o_1 \approx -0.06$, thus demonstrating co-existence of the tetrahedral and octahedral symmetries in the discussed nuclear ground-state configuration.

\subsection{Geometrical Properties of the Discussed Shapes}

Whereas the illustrations of the quadrupole shapes can frequently be found in the literature, the exotic symmetry examples are much less frequent. It will be instructive to show the corresponding competing shapes predicted by the total energy calculations shown in Fig.~\ref{Fig_12}.
\begin{figure}[h!]
\begin{center}
\includegraphics[width=0.45\columnwidth]{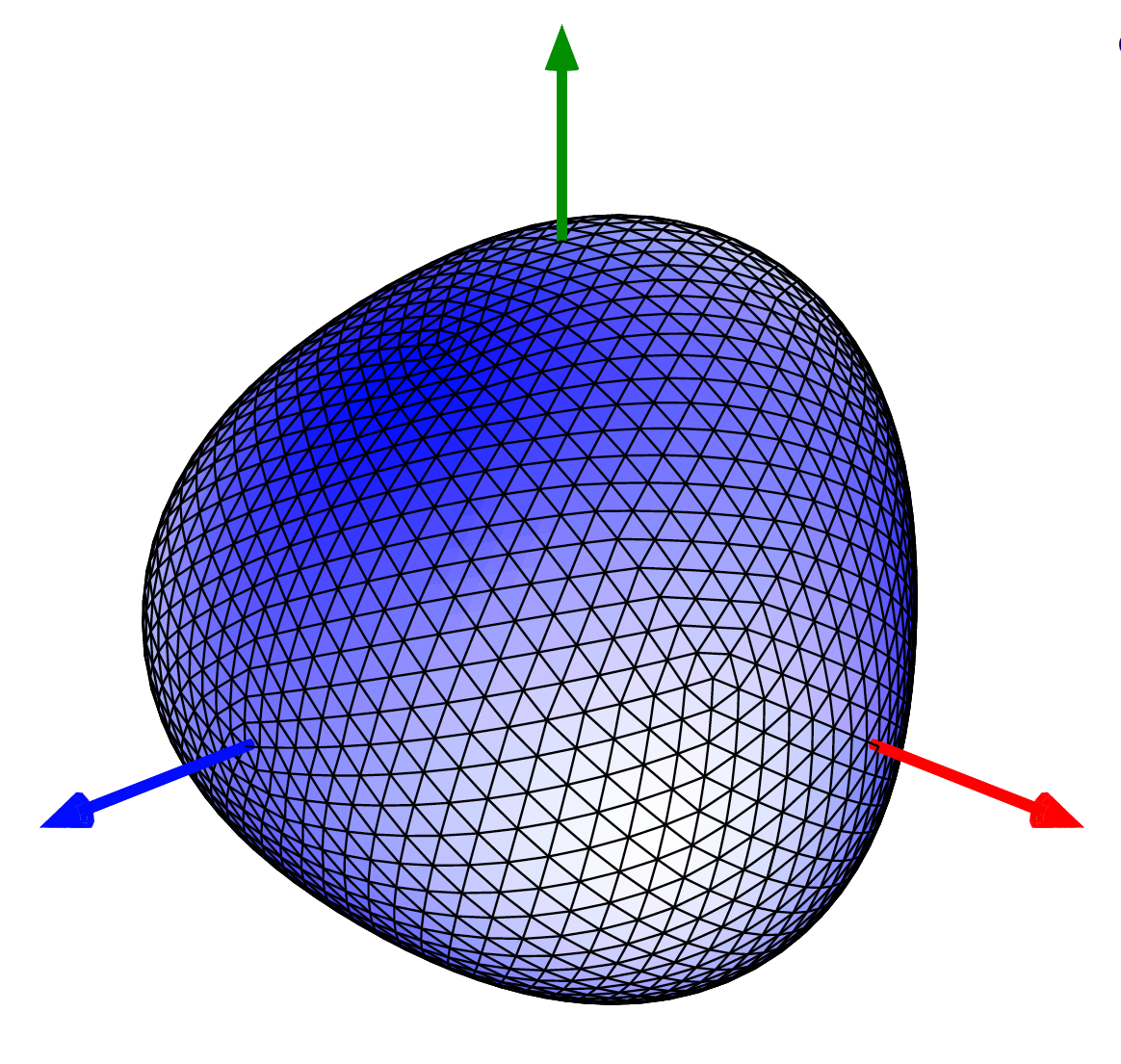}
\includegraphics[width=0.45\columnwidth]{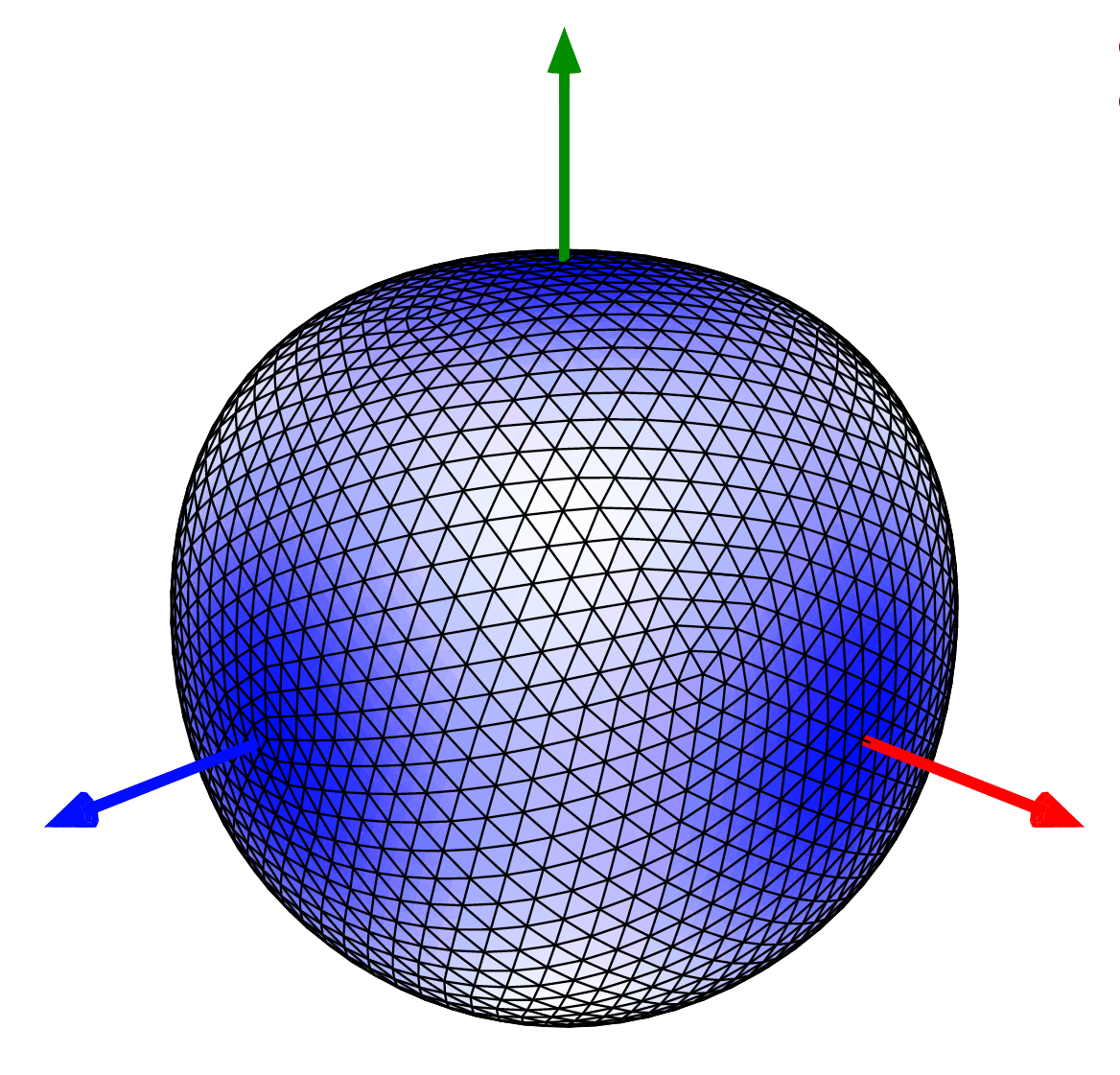}
\caption{Pedagogical illustrations of the predicted tetrahedral ($t_1=0.12,\,o_1=0$, left) compared to the octahedral ($o_1=-0.06, \,t_1=0$ right) competing shape configurations related to the potential energy minima visible in the preceding Figure. 
}
                                                                      \label{Fig_13}
\end{center}
\end{figure}
The resulting nuclear surfaces are presented in Fig.~\ref{Fig_13} showing examples of the shapes of pure tetrahedral (left) and octahedral (right) symmetries.

Let us remark that whereas the presented diagrams can be seen as an artist view of the ``pyramid like'' shapes (except that unlike Egyptian pyramids the ones discussed here have a triangular basis) as well as a ``diamond like'' shapes, the corresponding surfaces simulate the nuclear density distributions and do not show sharp edges or corners. These diagrams help imagining the discussed shapes and yet have a number of disadvantages, such as predefining the view angles and thus making it impossible for the reader to compare these geometric shapes seen from different view points.

It can sometimes be considered as a disadvantage of the discussed graphical representation that, especially for moderate deformations, it is not easy for a human eye to determine the degree of distortion of the spherical shape of the same volume in trying to answer the question how big, even in relative terms, are the discussed deformations and/or how strongly perturbing the spherical symmetry. The following diagrams were constructed to show the distances, expressed in femtometres, between the compared surfaces and the equivalent (reference) spherical surfaces seen as the functions of spherical angles $\vartheta$ and $\varphi$. Such a representation allows the reader to appreciate the symmetry of placements of characteristic points of the surfaces such as the angular positions of the minimum and the maximum distance points.
\begin{figure}[h!]
\begin{center}
\includegraphics[width=\columnwidth]{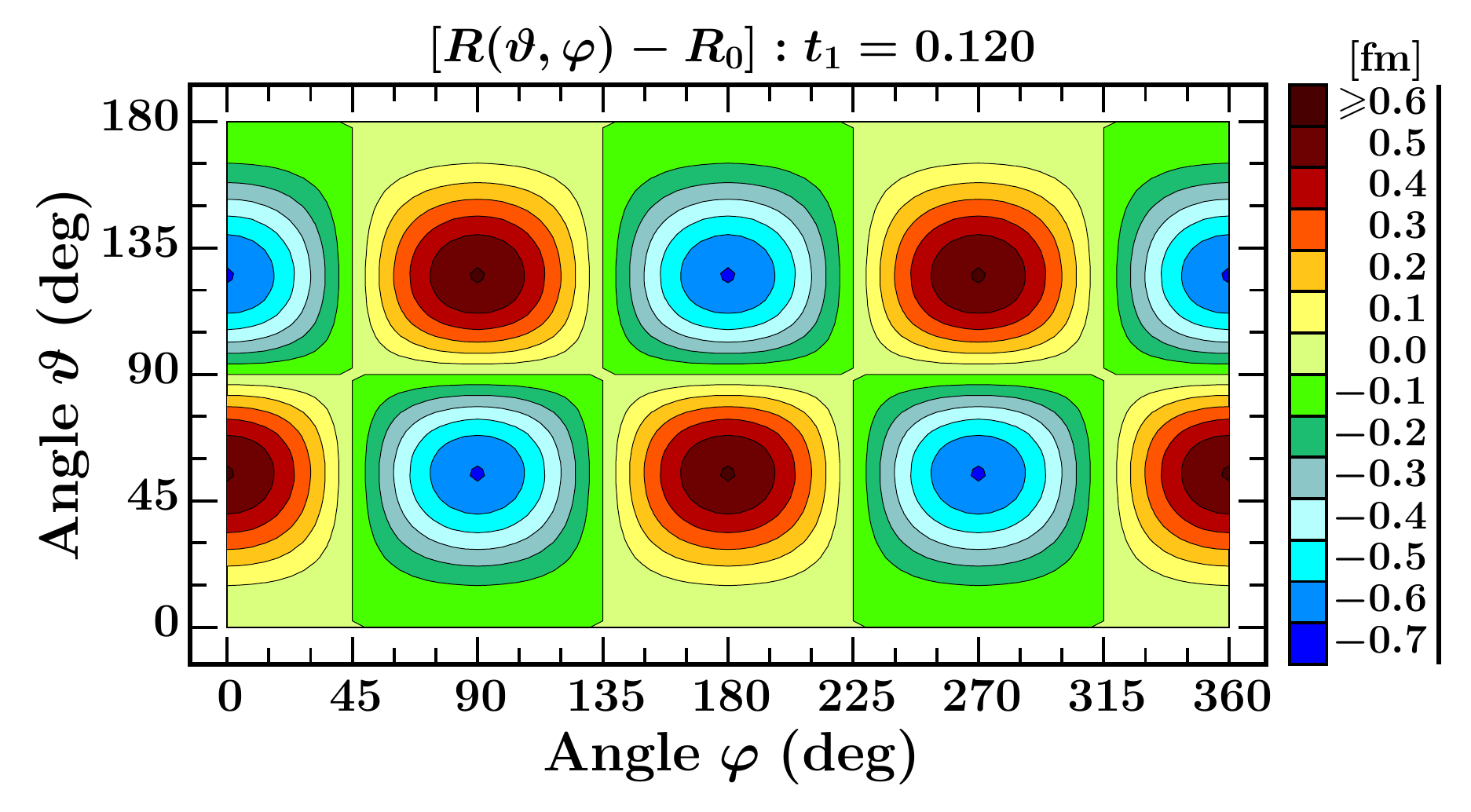}
\caption{Alternative (with respect to a geometrical surface) illustration for tetrahedral symmetry surface with $t_1=0.12$, Fig.~\ref{Fig_13}, left diagram. Function $R(\vartheta,\varphi)$ represents distances of points of nuclear surface from the origin of the reference frame, here displayed relative to the equivalent sphere of the radius of $R_o=6.4$\,fm for $^{152}$Sm. Observe repetitive positions of the green maxima on the contour plot corresponding to the relatively bright areas (`maxima') on the surface in Fig.~\ref{Fig_13}, left panel, similarly the repetitive positions of blue minima corresponding to the dark blue flat zones on the same `pyramid'.
}
                                                                      \label{Fig_14}
\end{center}
\end{figure}

%\vspace{-15mm}
\begin{figure}[h!]
\begin{center}
\includegraphics[width=\columnwidth]{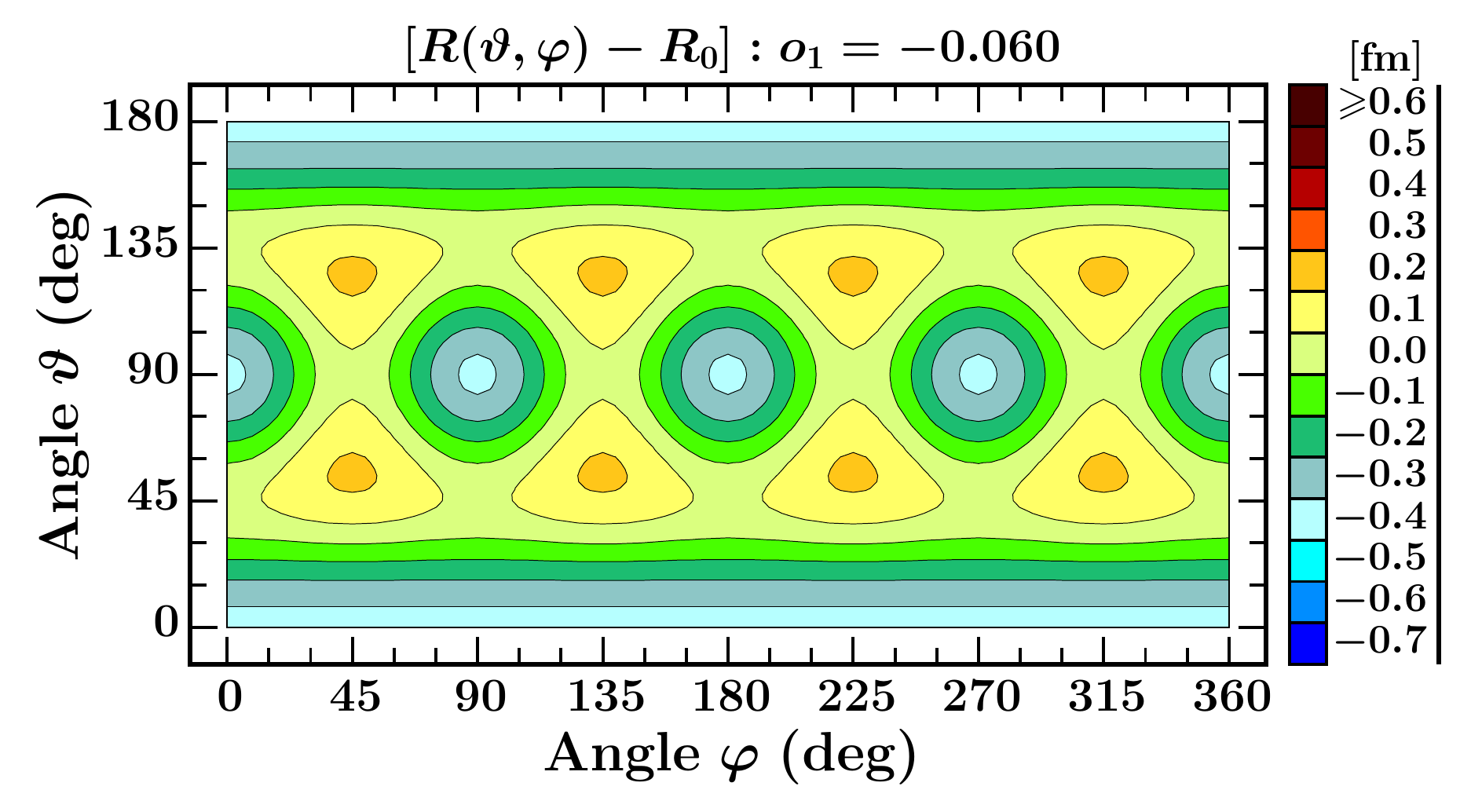}
\caption{Similar to the preceding one but for octahedral symmetry surface at $o_1=-0.06$. Observe the repetitive presence of the ``peaks'' (8 green zones corresponding to the relatively bright areas on the surface in Fig.~\ref{Fig_13}, right) and the flat areas (darker-blue zones  in Fig.~\ref{Fig_13}) -- at the angular distances of $\Delta \varphi = 90^o$ corresponding to $\vartheta=90^o$. The half-traced minima at $\varphi=0^o$ and $\varphi=360^o$ should be considered as one and the same. 
}
                                                                      \label{Fig_15}
\end{center}
\end{figure}

Illustrations in Figs.~\ref{Fig_14} and \ref{Fig_15}, show angular positions of the symmetry characteristics of the surfaces in Fig.~\ref{Fig_13} by focusing on the surface ``leading elements'' such as relatively flat zones (dark-blue areas in  Fig.~\ref{Fig_13}) and their exact correspondence to angular positions, compared with the positions of the ``peaks'' corresponding to the relatively brighter color zones in  Fig.~\ref{Fig_13}.

\begin{figure}[h!]
\begin{center}

\includegraphics[width=\columnwidth]{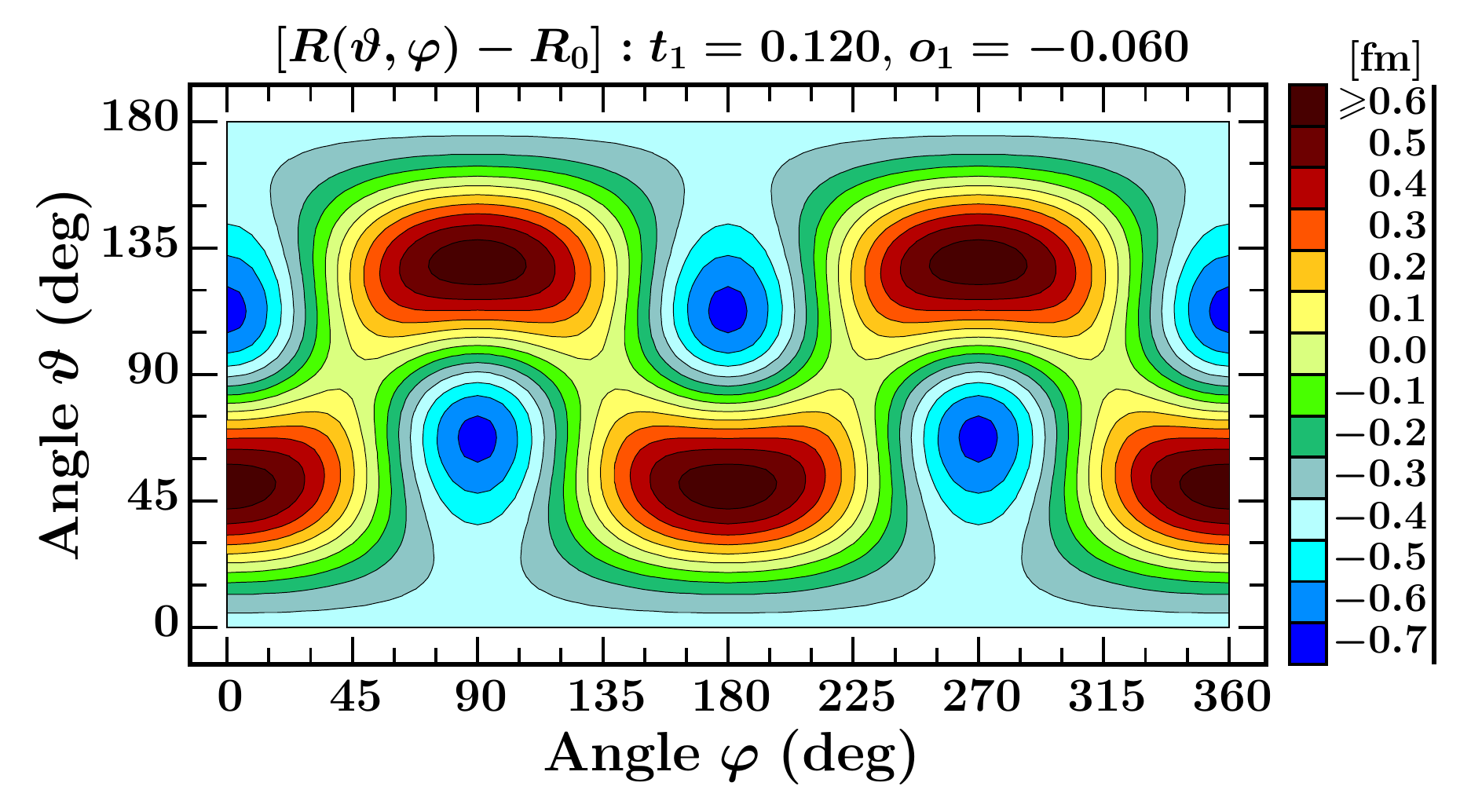}

\caption{Illustration analogous to the ones in the two preceding Figures, here representing precisely the shape at the local potential energy minima of Fig.~\ref{Fig_12}, i.e. combining the tetrahedral ($t_1=0.12$) and octahedral ($o_1=-0.06$) deformations at the same time: In a way ``summing up'' the information coming from the two competing shapes. Observe the relative similarities, when comparing this surface with the tetrahedral surface, Fig.~\ref{Fig_14}, of the characteristic distortions of the green zones, which remain centered at the original position points as well as the blue zones, which are getting distorted and displaced preserving overall similarities.
}
                                                                      \label{Fig_16}
\end{center}
\end{figure}

Please note characteristic ``topological'' differences between the diagram in Fig.~\ref{Fig_14} (showing 4 ``corners of the pyramid'' represented by 4 green ovals)
and the one in Fig.~\ref{Fig_15} (showing 8 ``corners of the diamond'' represented by 8 triangular forms). Finally let us observe that the extreme displacements from the sphere, which characterize the discussed surfaces in terms of the distance function visible from Fig.~\ref{Fig_13} correspond, for the tetrahedral form as an example,  to merely ($\pm 0.7$ fm), thus approximately, $\pm 10\%$ of the reference value of $R_o$. %Angle $\vartheta=90^o$ corresponds to $\mathcal{O}_{xy}$-plane.

\subsection{Identification of Tetrahedral Bands: Issues~and~Solutions}

In this section we will discuss quantum mechanisms ruling the  rotation of nuclei with tetrahedral symmetry -- focusing on the application of  the term ``bands'', which as it turns out may cause conceptual issues at the level of the very definition. Our search for the experimental evidence will rely on the use of Ref.~\cite{ensdf}. In fact, discussion which follows is partially provoked by diagrams present in certain data bases according to which any set of levels arranged as a vertical sequence (``ladder plot'') could be called a band -- without defining any choice criteria or preliminary specifications. 

In this context, let us pose the following problem: We are looking for experimental identification of the energy spectra, which satisfy the group-theory deduced spin-parity criteria collected in Eqs.~(\ref{Eqn_06}-\ref{Eqn_09}), referred to as tetrahedral/octahedral {\em bands} in specialized articles. Let us emphasize at this point that the group theory criteria in question are formulated in terms of restrictions on spins, parity of spins and parity of states; the only limitations concerning the energies are related to the degeneracies of certain levels. In other words no information about $E_I$ dependence on $I$.  When using the Ref.~\cite{ensdf} we must take into account that the usually strongest collective electromagnetic transitions (E1 and E2) do not populate neither depopulate the sought states because the corresponding reduced transition probabilities vanish in the tetrahedral/octahedral symmetry cases. Therefore the sought states are expected to be present in distinct i.e.,~separate diagrams illustrating various spectra of the data base. In a general search of this kind employing experimental data sets we will need to select for our purposes the level energies one by one, so that at the end we are left with a sequence of numbers labelled, following Eq.~(\ref{Eqn_06}), as 
\begin{equation}
E_{0^+}, E_{3^-}, E_{4^+}, E_{6^+}\approx E_{6^-}, E_{7^-}, E_{8^+}, E_{9+} \approx E_{9-},  \ldots
                                                                  \label{Eqn_11}
\end{equation}

Which arguments can be used to decide that the set of numbers obtained in this way represents experimentally identified tetrahedral {\em band\,}? We will discuss this issue in some detail in the following section.

\subsubsection{Notion of Bands: From Concepts to Nomenclature}

It will be instructive to recall the historical discussions introducing the notion of {\em bands} following up pioneering works of A.~Bohr and B.~Mottelson in the middle of the previous century. These authors, cf.~Ref.~\cite{ABM75}, often use rather general names such as collectivity or collective rotation as well as rotational energy or rotational spectra, i.e.,~not speaking directly about rotational bands -- the later terminology arriving later in the context of rotation of a symmetric top (a classical body with two of the three moments of inertia equal, also referred to as an axially symmetric rotor). Denoting the symmetry axis of the corresponding classical rotor $\mathcal{O}_3$ and the corresponding moment of inertia $\mathcal{J}_3\equiv \mathcal{J}_\parallel$, whereas the remaining ones  $\mathcal{J}_1=\mathcal{J}_2=\mathcal{J}_\perp$, after quantization of the related classical energy expression we find the corresponding operator, second order Hamiltonian, in the form
\begin{equation}
   \hat{H}_{\rm rot}^{(2)} = \frac{\hbar^2\hat{I}^2}{2\mathcal{J}_\perp}
   +\left[
         \frac{\hbar^2}{2\mathcal{J}_\parallel} - \frac{\hbar^2}{2 \mathcal{J}_\perp}
    \right] \hat{I}_3^2, \;\; \hat{I}^2=\hat{I}_x^2+\hat{I}_y^2+\hat{I}_z^2.
                                                                  \label{Eqn_12}
\end{equation}
The axial symmetry condition implies two consequences: Firstly, the angular momentum projection $K$ on the symmetry axis, is a constant of the motion with the condition $I \ge K$ and, secondly, there are no collective rotations about the symmetry axis.

The lowest energy states are then characterized by the condition $I=K=0$, and the rotational spectrum takes the form
\begin{equation}
   E_I = \frac{\hbar^2}{2 \mathcal{J}_\perp} I(I+1).
                                                                   \label{Eqn_13}
\end{equation}
Soon, the implied rotational energies proportional to $I(I+1)$ were confirmed experimentally, cf.~the historical articles, Refs.~\cite{ABM53} and \cite{FAP53}, completing, for each $K$, the angular momentum sequences entering Eq.~(\ref{Eqn_11}) as: 
\begin{equation}
   I=K,\,K+2,\,K+4,\, K+6, \, K+8, \, \rm etc.
                                                                   \label{Eqn_14}
\end{equation}
Identification of the {\em unique} $K=0$ sequences built on many ground-states of even-even nuclei, each sequence parameterized with a single constant $(\mathcal{J}\leftrightarrow\mathcal{J}_\perp)$, was immediately seen as a considerable success. Very likely the experimental evidence of these unique $(E-{\rm vs.-I)}$ sequences encouraged specific names for them: ground-state {\em bands} with the lowest energies called {\em band-heads}. From now on, the ``common sense'' definition of a {\em band} became that of a parabolic sequence induced by Eq.~(\ref{Eqn_12}).

%At this point we arrive at (as it may look) the clearly and uniquely defined traditional notion of the band which is represented by a parabolic energy-vs.-spin function in Eq.~(\ref{Eqn_12}), depending on a single parameter $\mathcal{J}\leftrightarrow \mathcal{J}_\perp$ which was interpreted as the nuclear moment of inertia. 

The concept of uniqueness evolved further with new advances in symmetry investigations. As it turned out, for non-axial rotors, the rotational spectra may contain more than one state at any given spin $I$, and, moreover, when including higher and higher angular momenta, the unique role of the ground-state moment of inertia became questionable since it turned out that expressions of the type
\begin{equation}
   E_I = A\,I(I+1) + B\,I^2(I+1^2) + C\, I^3 (I+1)^3 + \ldots
                                                                   \label{Eqn_15}
\end{equation}
can be empirically more successful. The later relation implies that the unique parabolic relations resulting from the original simplicity of the symmetric top Hamiltonian
\begin{equation}
   \hat{H}_{\rm rot}^{(2)} = \frac{\hat{I}^2_1}{2 \mathcal{J}_1}
                     + \frac{\hat{I}^2_2}{2 \mathcal{J}_2}
                     + \frac{\hat{I}^2_3}{2 \mathcal{J}_3}
                     \to \frac{\hat{I}^2}{2\mathcal{J}_\perp}
   +\left[
         \frac{1}{2\mathcal{J}_\parallel} - \frac{1}{2 \mathcal{J}_\perp}
    \right] \hat{I}_3^2
                                                                  \label{Eqn_16}
\end{equation}
can be seen as oversimplified and should be replaced by some generalized solutions. One possible way of such a generalization consists in replacing originally constant moments of inertia by angular-momentum dependent expressions constructed after certain modeling arguments. Another generalization is based on the idea of angular momentum dependence more involved as compared to the original quadratic form
\begin{equation}
   \hat{H}_{\rm rot}^{(2)} \to \hat{H}_{\rm rot}
   =\hat{H}_{\rm rot}^{(2)} + \hat{h}(\{p\},\hat{I_1},\hat{I_2},\hat{I_3}),
                                                                  \label{Eqn_17}
\end{equation}
where $\hat{h}$ may contain linear and/or higher than quadratic dependence on angular  momentum components and $\{p\}$ denotes a set of adjustable parameters. This second term in Eq.~(\ref{Eqn_17}) could be used to describe e.g,.~small, corrective deviations in the energy expression in Eq.~(\ref{Eqn_15}) from the simple parabolic E-vs.-I dependence. 

One can easily show that $\hat{H}_{\rm rot}^{(2)}$ representing quadratic dependence on angular momentum with all the three moments of inertia different is symmetric with respect to the point group D$_2$. In fact, the authors of Ref.~\cite{JDu01} introduce such a dependence in order to be able to model various, nontrivial exotic point group symmetries of the rotors with the help of specially constructed basis of tensors built out of angular-momentum operators. As indicated already in Ref.~\cite{ABM75}, Chapter 4, \S 4-5, considering rotors with more involved symmetries, for instance with two or more symmetry axes of order $n>2$ (as in the case of tetrahedral or octahedral symmetries) the tensor of inertia acquires spherical symmetry ($\mathcal{J}_1=\mathcal{J}_2=\mathcal{J}_3$) so that the operator $\hat{H}_{\rm rot}^{(2)}$ describes the system referred to as spherical top with the energy proportional directly to $I(I+1)$. Since there is no collective rotation possible, neither around the symmetry axis of the symmetric top nor any other axis of the spherical top, the concept of collective rotation in this case can be maintained thanks to the second term in Eq.~(\ref{Eqn_17}) -- or by changing strategy and working from the beginning with a microscopic theory Hamiltonian with a built in concept of rotation with respect to a laboratory reference frame (see below).

A few more comments about phenomenological rotor treatments will be of interest. To begin let us observe that in order to be able to determine the adjustable parameters $\{p\}$ in $\hat{h}$, a set of experimental data describing rotation of nuclei symmetric under a given point-group  must be found first. This, however, is impossible as long as the energy vs.~spin dependence remains unknown -- and this is exactly the case since group theory does not provide such a constraint.

Under these circumstances we turn to the discussion of suggestions which can be obtained from microscopic modeling approaches.

\subsubsection{Microscopic Description of Nuclear Rotation}

Starting research of nuclear tetrahedral symmetry has become possible thanks to predictions of the shell-structures in the form of tetrahedral symmetry induced gaps in single-nucleon spectra,  at proton and neutron numbers $Z_{\rm t},\,N_{\rm t} = 16, 20, 32, 40, 56, 68-70, 90-94, 112$  and $N_{\rm t}=136$ and 142, obtained originally with the Macroscopic-Microscopic Method (MMM), Refs.~\cite{XLi94,JDu88,JDu06}.  Sizes  of these gaps turned out to be comparable with the sizes of the traditional spherical-shell magic-gaps. Thanks to this information systematic predictions of nuclei with the potential energy minima indicating stable tetrahedral configurations became possible. The predictions of tetrahedral and `traditional' shape competition originally obtained using MMM turned out consistent with the numerous calculations using microscopic Skyrme-Hartree-Fock approaches, cf.~early~Refs.~\cite{STa98,MMa98,MYa01} and compare with the ones which followed, Refs.~\cite{POl06,KZb06,YSC08,KZb09}.

All these predictions addressed the occurrence of the static (no rotation) energy minima, ground-, or excited states, corresponding to the tetrahedral symmetry shapes. 

A break through has been achieved by Tagami and collaborators, who applied their Skyrme Hartree Fock Cranking approach with angular momentum and particle number projection techniques together with the tetrahedral-symmetry constraint to the realistic calculations of rotational properties of the selected doubly-magic tetrahedral nuclei; cf.~Ref.~\cite{STa12} for mathematical details and Ref.~\cite{STa13} for the realistic results.

Perhaps surprisingly, microscopic calculations of these authors reproduce the group theory results summarized by Eq.~(\ref{Eqn_06}) even though, as emphasized earlier in the Introduction, the computer programs used contain no information about group theory or symmetry considerations other that the shape-constraint $Q_{32}$. For reader's convenience the following diagram, adapted after Ref.~\cite{STa13}, illustrates the evolution with an increase of the tetrahedral constraint of both, the energy-vs.-spin dependence resembling more and more a parabola, as well as the arrival of characteristic degeneracies, cf.~also figure caption.

%Following these well known ideas will be helpful in constructing our methods of searching for exotic, beginning with tetrahedral symmetry. From our theory predictions we know that the tetrahedral deformation are relatively small. Illustrative cases in Sect. IVA (see also section IVC) indicate deviations from sphericity at the level of 10\% what suggest (without proving it) that the impact of $\hat{h}$ should be roughly 10 times smaller than that of the leading spherical top described by $\hat{H}_{\rm rot}^{(2)}$

%Let us pose the issue from the other side, in the way which will be needed in this article. Suppose a physicists found a sequence of nuclear excited states $E_I$ depending on nuclear angular momentum $I$. It will be natural to examine first of all the regularity of such a dependence plotting the corresponding energies as points within $(E,I)$ plane. What criterion should be applied to decide whether the so obtained sequence can be called a band?

\begin{figure}[ht!]
\begin{center}
\includegraphics[width=0.8\columnwidth]{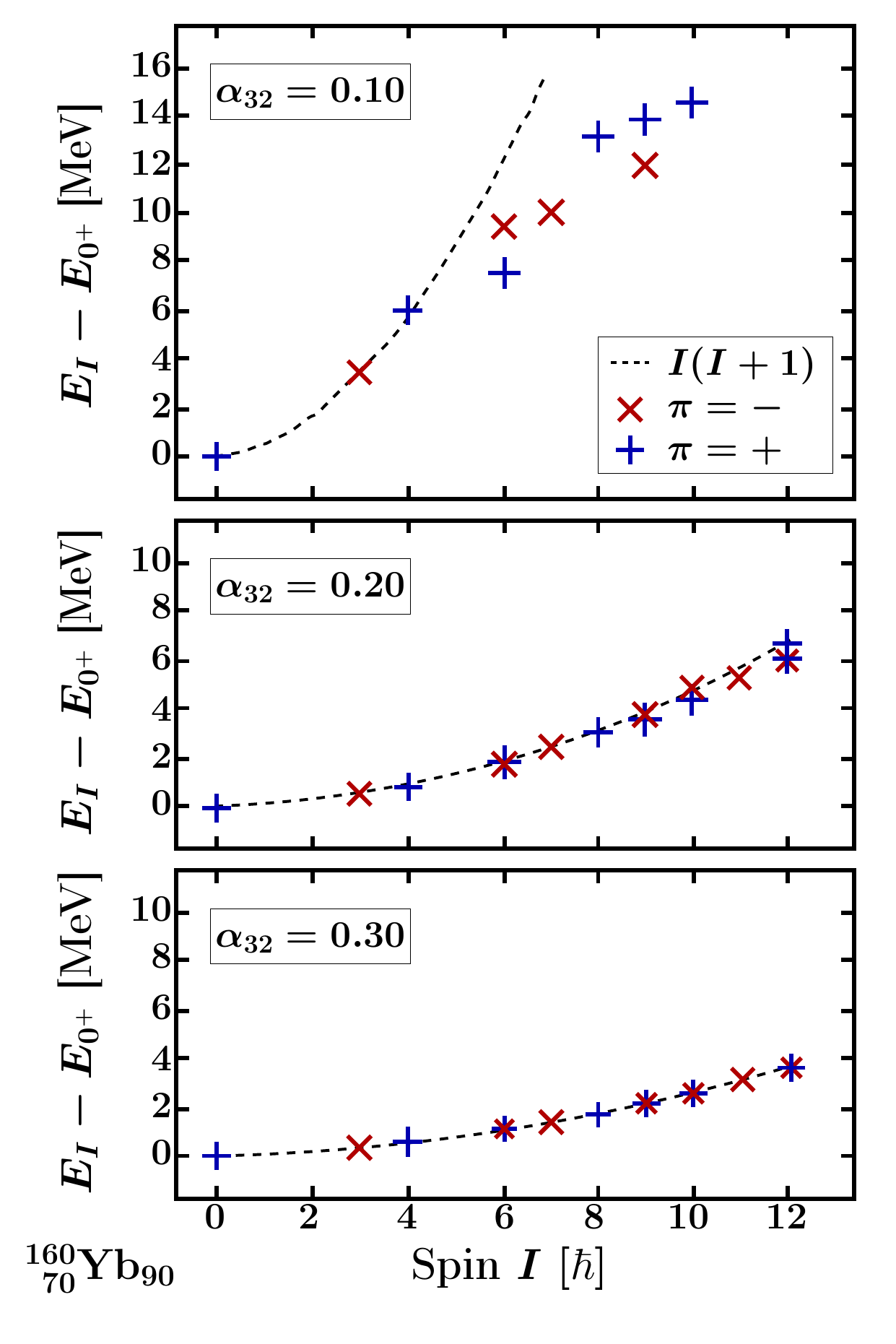}
\caption{Results of the angular-momentum and particle number projected Skyrme Hartree-Fock Cranking approach, which employs deformation constraint in the form of tetrahedral multipole moment $Q_{32}$. The diagram has been adapted after results of Ref.~\cite{STa13}. To facilitate discussion and comparisons with the rest of the article the drawings are labelled with the tetrahedral deformations $\alpha_{32}$ equivalent to the multipole moments used as constraints. Observe an increase of the tetrahedral symmetry impact following an increase of the corresponding deformation bringing the $E_I-{\rm vs.}-I$ energy dependence to a parabolic form -- whereas precision of the degeneracies at $I^\pi=6^\pm,9^\pm$ and $10^\pm$ (and even the triplet degeneracy predicted by group theory at $I=12$, two ``plus'' symbols overlapping with a red cross) increases visibly as well.
}
                                                                      \label{Fig_17}
\end{center}
\end{figure}

We use this information as the missing energy-vs.-spin criterion and test our measured $E_I$ states using a $\chi^2$ test.

\noindent
{\em Hartree-Fock Symmetry Modeling: Partial Conclusions.}\\[2mm] 
1) At sufficiently strong tetrahedral symmetry constraint the characteristic spectral features, including the T$_{\rm d}$ degeneracy pattern are perfectly reproduced providing the awaited information about the energy-vs.-spin (to a very good approximation parabolic) dependence;\\ 
2) At a decreasing constraint (decreasing T$_{\rm d}$ symmetry impact) the parabolic $(E-{\rm vs.}-I)$ dependence becomes only approximate; moreover, the projected HF theory provides less and less precise information about deviations from the smooth (parabolic) dependence. \\
3) As discussed in the next sections, our microscopic calculations predict deformations smaller than $\alpha_{32}\approx 0.2$, so that some deviations from the illustrated strong symmetry limits are to be expected. Knowing an experimental spectrum in the form of the hypothetic $E_I$ band we are in a position of estimating the expected position of the sought parabolic dependence by a $\chi^2$-adjustment and the level of uncertainty by calculating r.m.s.~deviations from the parabolic average.

\subsubsection{Tetrahedral Symmetry: Bands or Sequences}

\begin{figure}[ht!]
\begin{center}
\includegraphics[width=\columnwidth]{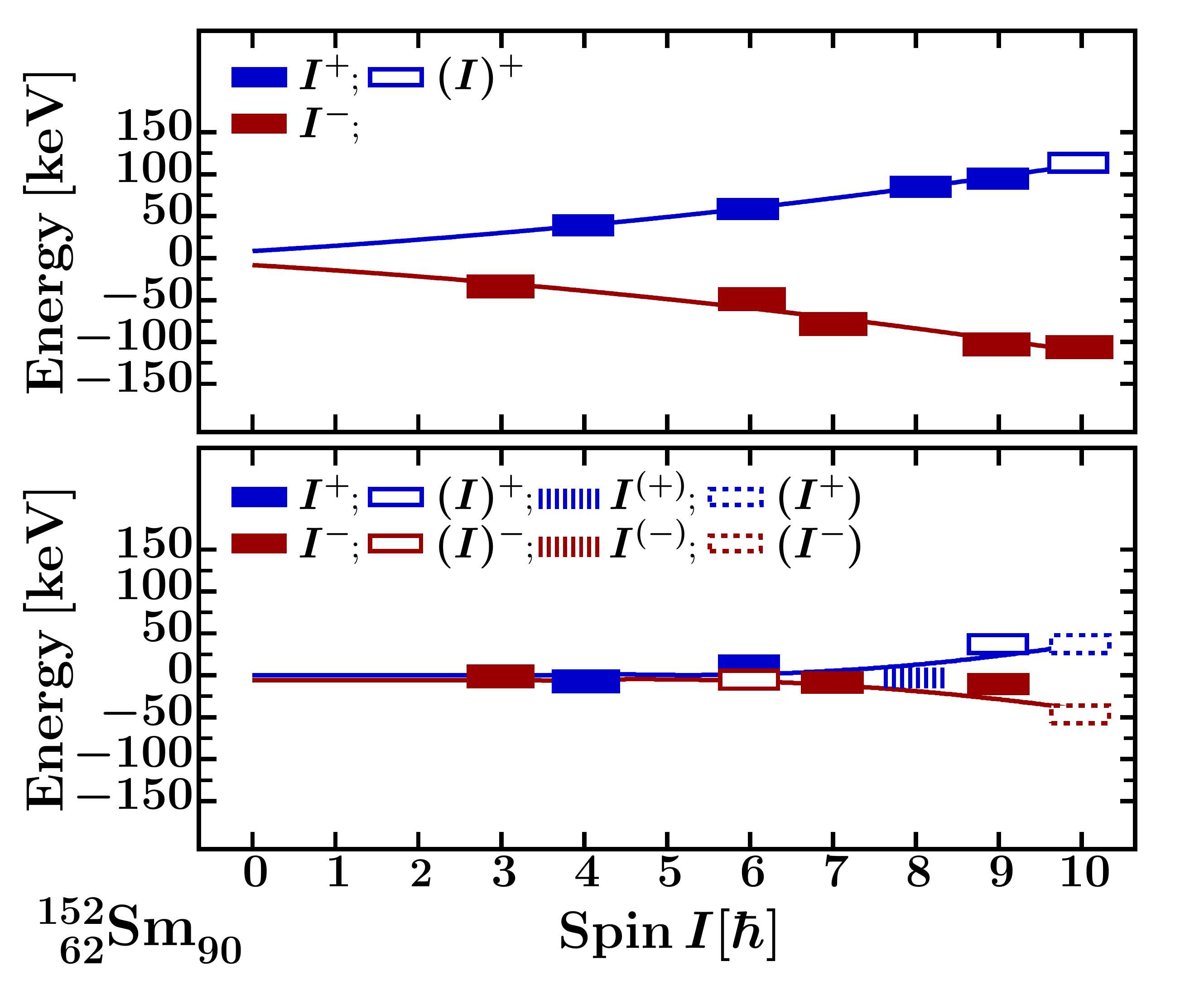}
\caption{Opposite parity branches of experimental T$_{\rm d}{(1)}$ band, top, and T$_{\rm d}{(2)}$ band, bottom. Reference parabolas of both bands were subtracted  to make small differences better visible. Near degeneracy of both branches of T$_{\rm d}{(2)}$,  manifests nearly exact tetrahedral symmetry condition, Eq.~(\ref{Eqn_06}), as opposed to a significant deviation for T$_{\rm d}{(1)}$, interpreted as relatively strong tetrahedral symmetry breaking in T$_{\rm d}{(1)}$ caused by the impact of the octahedral one. The spin-parity uncertainties marked with the help of parentheses are indicated in the legend. 
}
                                                                      \label{Fig_18}
\end{center}
\end{figure}

Two, approximately parabolic sequences of what we call T$_{\rm d}{(2)}$ band, observed in this article, are shown in Fig.~\ref{Fig_18}, bottom panel. They correspond to the opposite parities and were obtained by fitting parabolic energy-vs.-spin dependence $E_I=a*I^2 +b*I +c$. Both sequences meet at the common unobserved level interpreted as $I^\pi=0^+$ band-head. The position of the latter was estimated {\em via} extrapolation of the two parabolas down to $I=0$. 

Analogous results for the T$_{\rm d}{(1)}$ band, from Ref.~[8], are shown in the top panel of the Figure. We find the two band-head energies at $E_{\rm b.h.}[T_{\rm d}(1)]=1.396$\,MeV and $E_{\rm b.h.}[T_{\rm d}(2)]=1.666$\,MeV. They are interpreted as tetrahedral vibrational excitations, cf.~Appendix in Ref.~\cite{JDEPJ}, in which discussion overviewing the long standing issue of vibrational band-heads in, among others, alpha-cluster modeling of nuclear structures can be found.

We proceed comparing the characteristic features of the T$_{\rm d}{(1)}$ band and the newly identified T$_{\rm d}{(2)}$ band. Both, satisfy the parabolic relation of Eq.~(\ref{Eqn_06}), yet with strongly differing root-mean-square deviations. From parabolic fits we find: $\rm rms[T_d{(1)}]=81.3$\,keV against $\rm rms[T_d{(2)}]=23.4$\,keV. Analogous deviations for opposite parity branches are smaller: $\rm rms[A_{\rm 1g}{(1)}]=1.6$\,keV, $\rm rms[A_{\rm 1g}{(2)}]=11.4$\,keV, $\rm rms[A_{2u}{(1)}]=7.5$\,keV and $\rm rms[A_{2u}{(2)}]=7.6$\,keV.

%{\color{red} Irene: What are the values of $\mathcal{J}_{A_{1g}}$ and $\mathcal{J}_{A_{2u}}$}
Because of the splitting of the two parabolas in the case of T$_{\rm d}(1)$ the implied effective moments of inertia  $\mathcal{J}_{A_{1g}}$ and $\mathcal{J}_{A_{2u}}$  differ. We interpret the so obtained significant differences between them by saying that $T_{\rm d}{(1)}$ manifests O$_{\rm h}$ symmetry since effective moments of inertia of the two branches visibly differ in accordance with the liberty offered by the group theory predictions in this case, whereas in $T_{\rm d}{(2)}$ case these moments are nearly the same what signifies domination of tetrahedral symmetry with two opposite parity branches nearly degenerate as expressed by Eq.~(\ref{Eqn_06}). 

We say that tetrahedral symmetry is spontaneously broken by octahedral one in the $T_{\rm d}{(1)}$ case,  whereas it is fully present in the $T_{\rm d}{(2)}$ case. 

%%%%%%%%%%%%%%%%%%%%%%%%%%%%%%%%%%%%%%%%%%%%%%%%%%%%%%%%%%%%%%%%%%%%%%%%%%%%%%%%%%%%%

\subsection{Bohr Theory of Collective Motion in Tracing Exotic Symmetries}

We intend to combine nuclear mean field theory and Bohr nuclear collective motion theory for interpretation of exotic symmetries and their competition discussed so far.
We will examine collective nuclear motion by employing the realization of the  Bohr approach in Ref.~\cite{DRo19}. It involves deformation dependent nuclear inertia tensor $B_{nm}$, where indices $n$ and $m$ enumerate deformations. Components of  $B_{nm}$ are calculated within mean-field approach and a new formulation of the nuclear adiabaticity concept. Bohr Hamiltonian of the form
\begin{equation}
   \hat H
   =
   - \frac{\hbar^2}2 \hat\Delta + V(q) ,
                                                                     \label{Eqn_18}
\end{equation}
involves Laplace operator, $\hat\Delta$, here in 2-dimensional space $\{q_1=t_1, q_2=o_1\}$ 
\begin{align}
   \hat\Delta 
   &=
   \sum_{m,n=1}^{d=2}
   \frac1{\sqrt{|B|}} \frac\partial{\partial q^n}
   \bigg(\sqrt{|B|} B^{nm} \frac{\partial \;\;}{\partial q^m}\bigg) ,
                                                                     \label{Eqn_19}
\end{align}
where $|B|$ denotes determinant of the inertia tensor. Collective Schr\"odinger equation reads
\begin{equation}
   \hat H\Psi_{i} 
   = \mathcal{E}_{i}\Psi_{i} ,
                                                                    \label{Eqn_20}
\end{equation}
with $\mathcal{E}_i$ for $i=1,2,3,$ ...  denoting energies of the sought collective vibrational spectrum.

We are going to calculate probability density functions, $\propto\Psi^*\Psi$, which allow to define the quantum probability of finding the system within volume $dV$ in the deformation space:
\begin{equation}
 d \mathcal P(q) 
 \stackrel{df.}{=} 
 \Psi^*_i(q) \Psi_i(q) \sqrt{|B|}\,dV  ,\quad dV \equiv dq_1 dq_2 .
                                                                     \label{Eqn_21}
\end{equation}
Knowing this quantity is essential in avoiding interpretation of static potential energy minima on flat surfaces as the physical equilibrium comparable with experiment. The reasons are provided by Bohr theory, according to which a deformed nucleus is a quantum system moving in its deformation space and the physical meaning of each deformation point is given by density of probability of finding the nucleus in this point. 

Consequently, the physically meaningful information should be sought in probabilistic terms: Maxima of the probability density function suggesting the most probable deformations (comparable with experiment) and the probability density integrated over the competing deformation zones determining the physically meaningful shape competition. Observe that probability density is defined by the wave functions, $\Psi_i(q)$, solutions of the Schr\"odinger equation with the Bohr model Hamiltonian, Eq.~(\ref{Eqn_18}), in which the deformation-dependent potential with its local minima and the separating barriers contributes directly to the shape competition phenomena,  together with deformation-dependent collective inertia tensor. 

Qualitatively, the bigger the inertia of the system in a given deformation space, the bigger the probability of finding it in the corresponding zone since $d \mathcal P(q) \propto \sqrt{|B|}$. In other words, and in accordance with the classical physics intuition, the bigger the inertia the slower the classical motion.  

\subsection{Dynamic vs.~Static Equilibrium Deformations}

We have calculated the potential energy surfaces $V(q)$ and the components of the inertia tensor $B^{nm}(q)$, compare Eqs.~(\ref{Eqn_16}) and (\ref{Eqn_17}), and we solved the Schr\"odinger equation (\ref{Eqn_18}) representing the Bohr collective model. In the following we will compare the solutions for the lowest vibrational state energies $\mathcal{E}_1$, $\mathcal{E}_2$ and $\mathcal{E}_3$.  

\begin{figure}[h!]
\begin{center}
\includegraphics[width=\columnwidth]{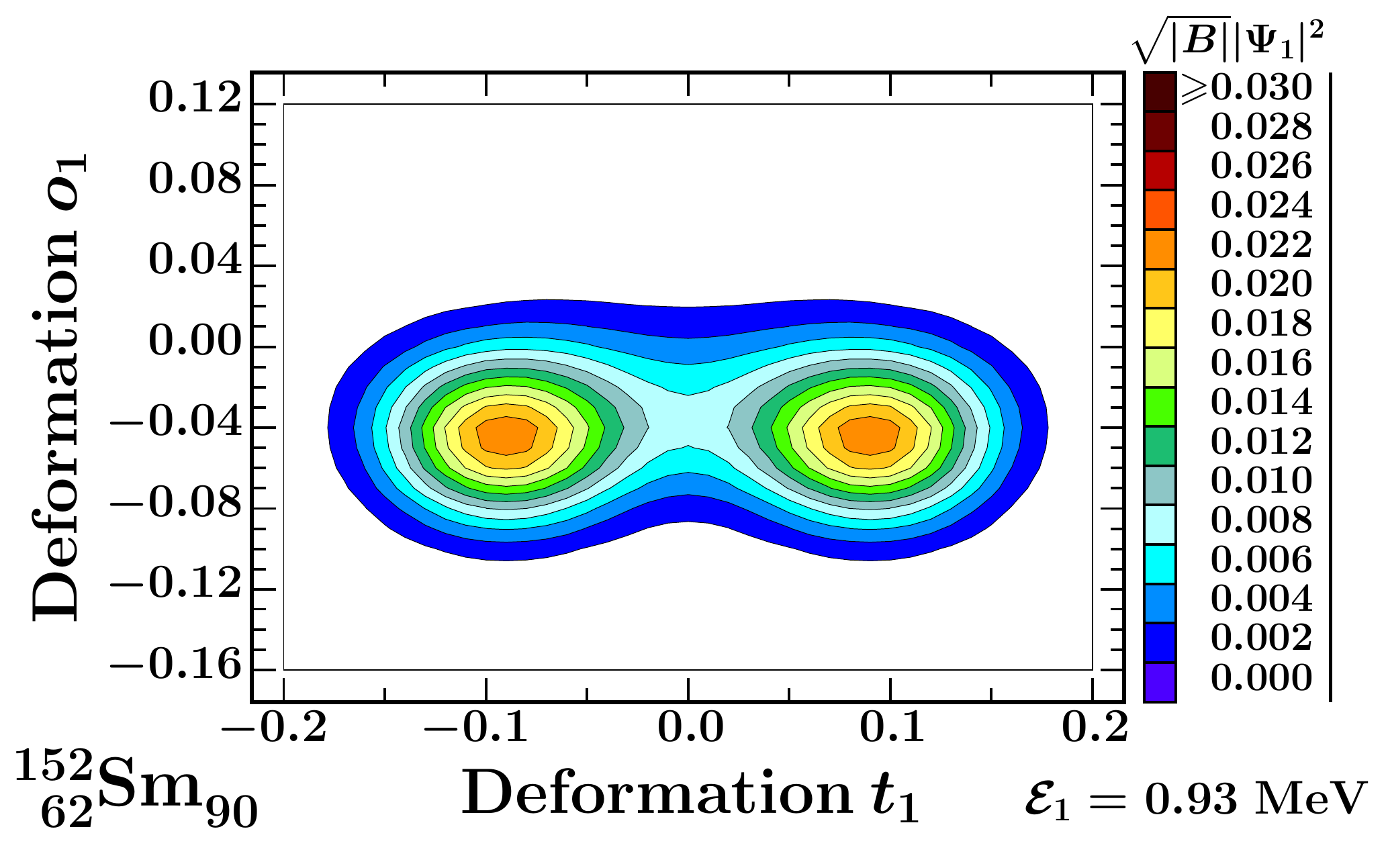}
\caption{Probability density distribution $\Psi^*_i(q) \Psi_i(q) \sqrt{|B|}$, cf.~Eq.~(\ref{Eqn_19}), for the lowest energy solution, $i=1$. Since the potential walls are steep and the barriers separating the energy minima are high -- the positions of the maxima of the probability distributions are very close to the positions of the static minima for the lowest energy vibration $\mathcal{E}_1=0.93$~MeV, compare with the contour plot in Fig.~\ref{Fig_12}.
}
                                                                      \label{Fig_19}
\end{center}
\end{figure}

\begin{figure}[h!]
\begin{center}
\includegraphics[width=\columnwidth]{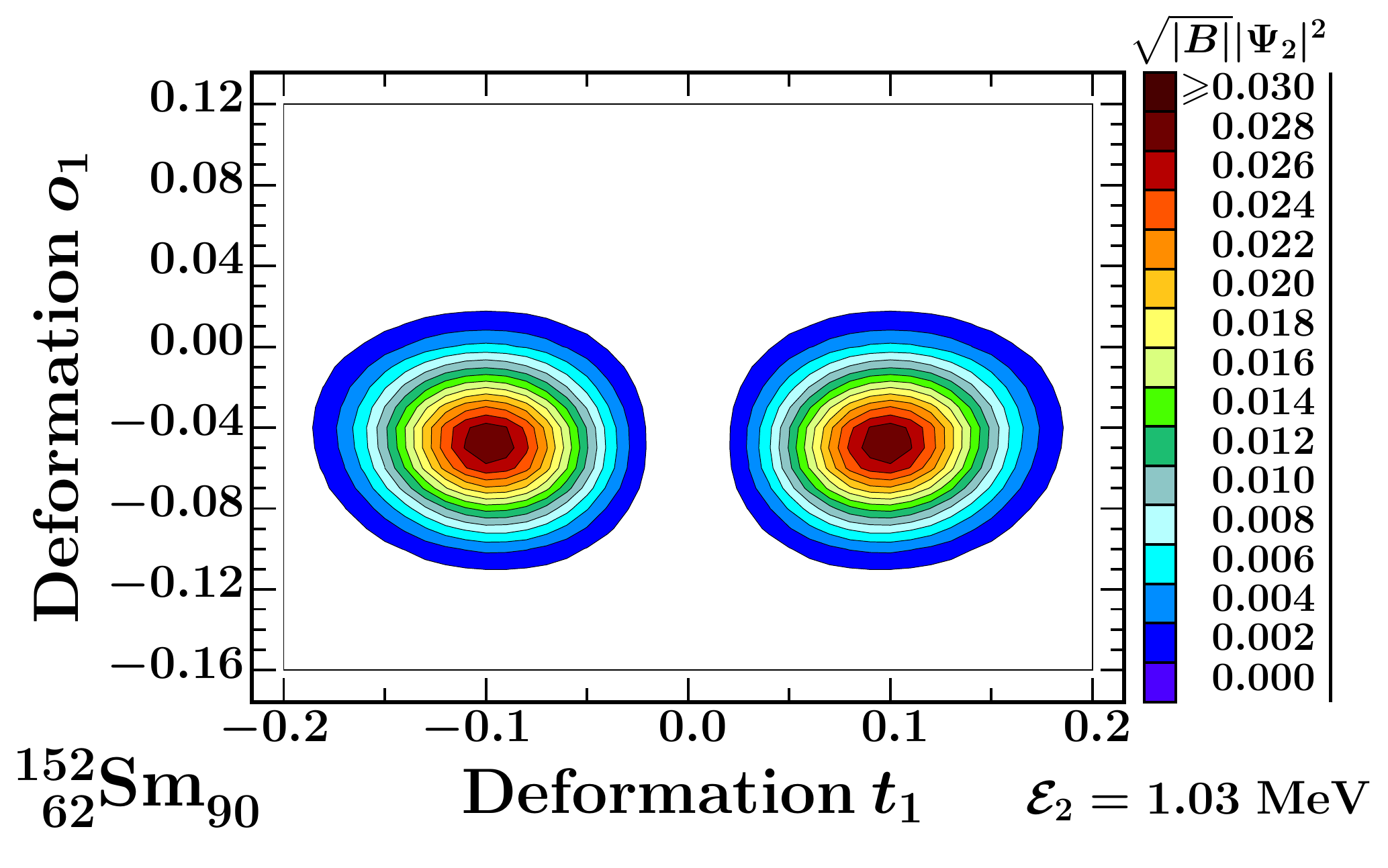}
\caption{Illustration analogous to the preceding one for the ``doublet solution'' $\mathcal{E}_2 \approx 1.03$\,MeV. In accordance with the comments given in the text all the properties are very close to the ones at vibration $\mathcal{E}_1=0.93$~MeV.
}
                                                                      \label{Fig_20}
\end{center}
\end{figure}

The potential barrier separating two minima in Fig.~\ref{Fig_12} is relatively high, $V_B\approx 2$\,MeV. One can easily show by semi-quantitative arguments using solutions of the Schr\"odinger equation with harmonic oscillator potential, that for sufficiently high effective inertia the two lowest energy solutions form approximately a doublet, whereas the wave functions present far going similarities. Our realistic solutions illustrated in Figs.~\ref{Fig_19} and \ref{Fig_20} confirm these expectations, both in terms of close energy values and very similar structures of the probability density functions with two maxima corresponding to tetrahedral deformations $t_1 \approx \pm 0.10$ coexisting with the octahedral one at $o_1 \approx -0.06$.

Our methods of modeling do not involve the reaction mechanism so that we are not in a position of arguing, which of the vibrational solutions will be populated by the reaction used in a discussed experiment. Instead we compare the nuclear structure predictions with the observed properties of the spectra and formulate what we consider the most direct interpretation.  

\begin{figure}[h!]
\begin{center}
\includegraphics[width=\columnwidth]{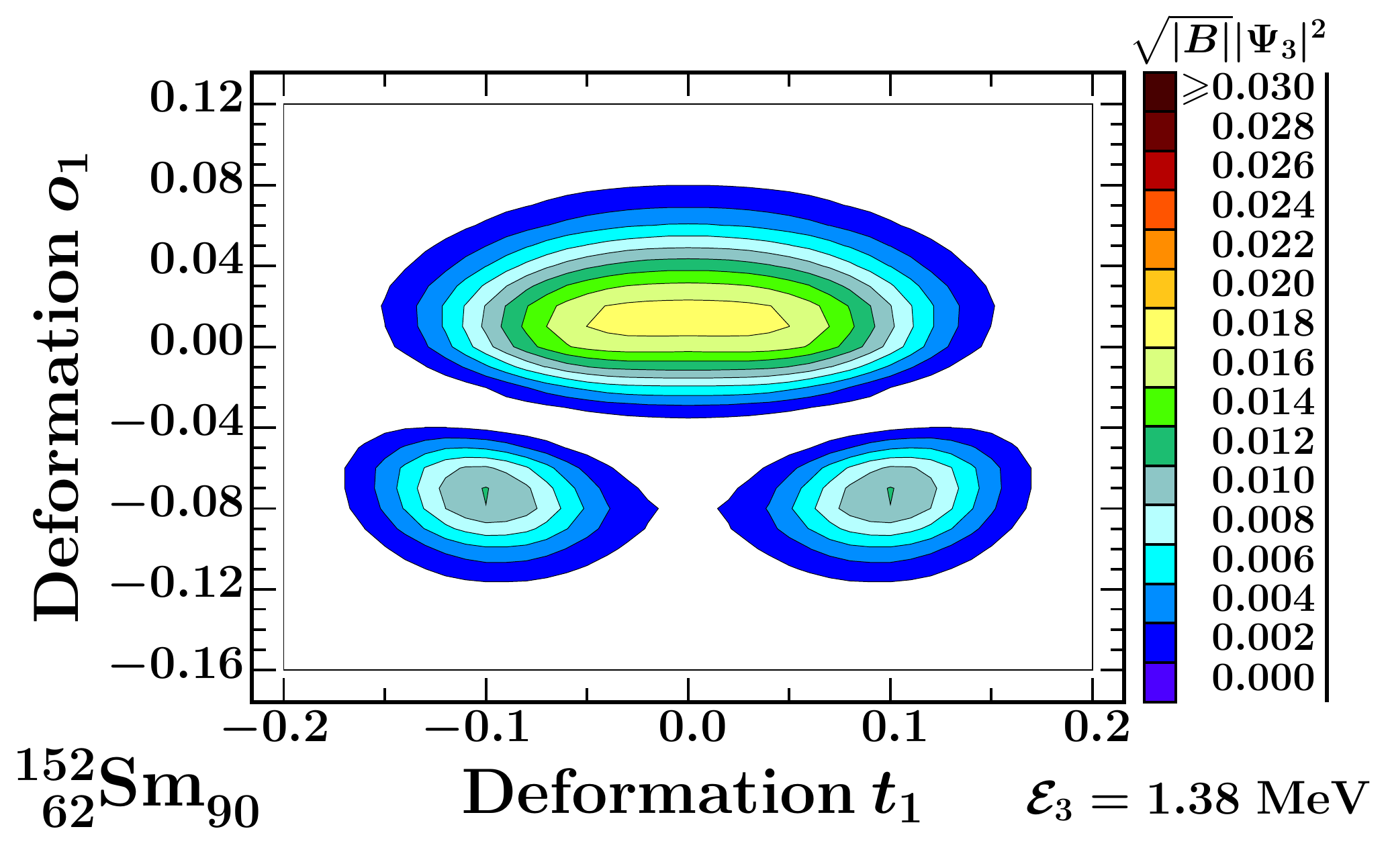}
\caption{Illustration analogous to the preceding ones but for the next excited state, $\mathcal{E}_3=1.38$~MeV. We have performed the probability integrations over the zone corresponding to the upper cloud with the result close to 75\% compared with the competing zone of the two symmetric smaller clouds giving the remaining 25\%\,. Thus it follows that the dominating structure of this state corresponds to vanishing octahedral component, $o_1 \approx 0$, confirming in this way the domination of tetrahedral symmetry discussed in detail in the text.
}
                                                                      \label{Fig_21}
\end{center}
\end{figure}

Results of our Bohr model calculations for the next vibrational state, $\mathcal{E}_3 \approx 1.38$\,MeV are shown in Fig.~\ref{Fig_21}. They suggest that the most probable deformation scheme corresponds to vanishing octahedral shape component and the domination of the tetrahedral one, thus in a qualitative agreement with the discussion of the ``spontaneous symmetry breaking'' presented in Sect.~IV B, see also caption of the Figure for more details.

We proceed calculating the most probable (dynamical) deformations taking as their measure $q_n^{\rm dyn} \equiv\sqrt{{\langle q_n^2 \rangle}}$, where 
\begin{equation}
 {\langle q_n^2 \rangle} 
 \stackrel{df.}{=} 
 \int \Psi^*_i(q) q_n^2\, \Psi_i(q) \sqrt{|B|}\,dV  ,\quad dV \equiv dq_1 dq_2 .
                                                                     \label{Eqn_22}
\end{equation}
Let us mention at this point that the usually employed simple expected value of the octupole deformation will better be avoided in this case because for odd-$\lambda$ deformations, like $t_1\equiv \alpha_{32}$, the potential satisfies the symmetry property
\begin{equation}
   V(-\alpha_{\lambda{\rm -odd},\mu})
   =
   V(+\alpha_{\lambda{\rm -odd},\mu}).
                                                                     \label{Eqn_23}
\end{equation}
It follows that for a constant inertia tensor the expected value of the discussed deformation vanishes identically
\begin{equation}
 {\langle q_n \rangle} 
 \stackrel{df.}{=} 
 \int \Psi^*_i(q) q_n\, \Psi_i(q) \sqrt{|B|}\,dV \equiv 0,\; \sqrt{|B|} =\rm constant ,
                                                                    \label{Eqn_24}
\end{equation}
so that in this case, the non-vanishing ${\langle q_n \rangle}$ measures more the asymmetry of the determinant than anything else.

Let us turn to the probability density distributions $(\sqrt{|B|}) |\Psi_i|^2$ for low energy solutions, $\Psi_i$ with $i=1$ and 3, illustrated in Figs.~\ref{Fig_19} and \ref{Fig_21}. 

For $\Psi_1$, after integrating, we find the dynamical equilibrium values 
\begin{equation}
   \Psi_1:\;\; t_1^{\rm dyn}\approx 0.09 \;\; {\rm and} \;\; o_1^{\rm dyn}\approx-0.04
                                                                     \label{Eqn_25}
\end{equation}
compared with the static results
\begin{equation}
   \Psi_1:\;\; t^{\rm stat}_1 = 0.12 \;\; {\rm and} \;\; o^{\rm stat}_1 \approx -0.06
                                                                     \label{Eqn_26}
\end{equation}

For $\Psi_3$ instead we find
\begin{equation}
   \Psi_3:\;\; t_1^{\rm dyn}\approx 0.07 \;\; {\rm and} \;\; o_1^{\rm dyn}\approx 0 .
                                                                     \label{Eqn_27}
\end{equation}
We conclude that for the vibrational solution $\Psi_1$ which we interpret as tetrahedral ground-state of the T$_{\rm d}{(1)}$ band we should admit the presence of coexisting T$_{\rm d}$ and O$_{\rm h}$ symmetries, with the implication of the spontaneous tetrahedral symmetry breaking by the octahedral one manifested by significant splitting of the positive and negative parity sequences visible in the experimental results in Fig.~\ref{Fig_18}, top panel. For the vibrational solution $\Psi_3$ which we interpret as the band-head of the tetrahedral band T$_{\rm d}{(2)}$ we find the octahedral dynamical equilibrium value approximately vanishing, $o^{\rm dyn}_1 \approx 0$, thus suggesting that tetrahedral symmetry T$_{\rm d}$ is the only one remaining present and providing the natural interpretation for the nearly coinciding opposite parity partners of T$_{\rm d}{(2)}$ band, in agreement with experimental results in Fig.~\ref{Fig_18}, bottom panel.\\

{Finally let us address some qualitative comments related to the observation of a dominance of the I $ \to$ I $^>$ type of transitions in the context of the spectroscopic properties of the T$_{\rm d}(2)$ band, as found in Section IV. We believe that this mechanism has to do with the breaking in the strongly non-axial tetrahedral and/or octahedral configurations  of the ``traditional'' axial-symmetry leading to the presence of the so-called K-bands, and conservation of the K quantum number discussed frequently in the literature.

Indeed, as it is well known from quantum mechanics, the absence of axial symmetry in a nucleus  causes mixing in terms of both angular momentum $\hat{\rm I}$  and its projection $\hat{\rm I}_3$ characteristics in the body-fixed reference frame with the result that the associated quantities are not conserved; the related observables are recovered within mean field theory using specifically designed projection techniques.

One may expect that such a mechanism introduces mixing of various angular momentum components via symmetry breaking interactions with the result that higher spin wave functions usually corresponding to higher excitation energy will mix with the lower energy solutions.

It then follows that the angular momentum projection techniques within the mean-field theory treatment could produce higher spin contributions at lower energies positively contributing to the discussed feeding competition. The projected mean-field theory treatment of this problem by-passes the limitations of the present project  and will be discussed elsewhere.
}

{\bf Conclusions.} The singles and coincidence analysis along with the determination of angular distribution coefficients, DCO and linear polarization in a dedicated experiment on the $^{152}$Sm nucleus indicate the presence of a new rotational band, referred to as T$_{\rm d}{(2)}$ to distinguish from the one identified earlier by other authors, denoted T$_{\rm d}{(1)}$, with properties resembling closely those foreseen by theory for tetrahedral symmetry bands. Even if we believe that the proposed interpretation is likely, the final confirmation of the T$_{\rm d}{(2)}$-band as tetrahedral symmetry band cannot be strongly claimed at this time because not all spins and parities could be uniquely identified from our experimental evidence. Hopefully, missing experimental information about  spins and parities will soon become available thanks to dedicated future experiments.

Group-theory arguments and microscopic mean-field and Bohr collective motion theory calculations suggest different realizations of the symmetries and symmetry breaking patterns in T$_{\rm d}{(1)}$ and T$_{\rm d}{(2)}$ bands, in agreement with actual experimental results and provide theory interpretations of specific degeneracies within T$_{\rm d}{(2)}$.

{\bf Acknowledgements.} Authors gratefully acknowledge the  efforts of the K-130 cyclotron operation team for providing good quality $\alpha$ beam. Support from the French-Polish collaboration COPIN is acknowledged.  The authors also acknowledge the efforts of those who assisted in setting up and maintaining the array.

\end{document}